\documentclass[letterpaper,twocolumn,10pt]{article}
\usepackage{usenix2019_v3}
\usepackage
[english]{babel}
\usepackage{blindtext}
\usepackage{amssymb}                 
\usepackage[varqu]{zi4}               
\providecommand{\Description}[1]{}   

\usepackage{adjustbox}
\usepackage{babel}
\usepackage{ifoddpage}
\usepackage{pifont}
\usepackage{siunitx}
\usepackage{multirow}
\usepackage{adjustbox}
\usepackage{epigraph}
\usepackage{tikz}

\usepackage{amsmath}
\usepackage{enumitem}
\usepackage[most]{tcolorbox}
\usepackage{multicol,lipsum}
\usepackage[utf8]{inputenc}
\usepackage{algorithmicx}
\usepackage{algorithm}
\usepackage{tabularx,booktabs}
\usepackage{tabularray}
\usepackage[noend]{algpseudocode}
\usepackage{soul}

\usepackage{appendix}
\usepackage{balance}
\usepackage{makecell}
\usepackage[skip=4pt]{caption}
\usepackage{diagbox}
\usepackage{float}
\usepackage{graphicx}
\usepackage{hyperref}
\usepackage{siunitx}
\usepackage{subcaption}
\usepackage{threeparttable}
\usepackage{url}

\usepackage{verbatim} 
\usepackage{xspace}
\usepackage{xcolor}
\usepackage{totpages}
\usepackage{mathtools}
\usepackage{physics}
\usepackage{array}
\usepackage{pdfpages}
\usepackage{adjustbox}
\usepackage{float}
\usepackage{environ}
\usepackage[normalem]{ulem}
\usepackage{hyperref}
            
\usepackage{tcolorbox}
\definecolor{mycolor}{rgb}{0.122, 0.435, 0.698}
\usepackage{xparse}

\def\noeditingmarks{}

\ifx\noeditingmarks\undefined
  
\else
  
\fi

\NewEnviron{commentbox}{%
  \vspace{0.2cm}
  \begin{tcolorbox}[
    width=.48\textwidth,
    colframe=mycolor,
    coltext=mycolor,
    colback=white!10!white,
    boxrule=0.5pt,
    arc=4pt,
    boxsep=0pt,
    left=6pt, right=6pt,
    top=6pt, bottom=6pt
  ]
  \BODY
  \end{tcolorbox}
}

\newcommand{\comments}[1]{}

\newcommand{\Para}[1]{\par\vskip 2pt plus 1pt\noindent\textbf{\textit{#1}}}
\newcommand{\sysname}[0]{\texttt{Stephen}\xspace}

\newcommand{\ag}[1]{}

\usepackage{listings}
\lstdefinestyle{rocqspec}{%
  basicstyle=\scriptsize\ttfamily,
  alsoletter=.,  
  keywords=[1]{bool,nat,loc,u64,u8,val,loghist,log_entry_type,byte_string,chan},
  keywordstyle=[1]\color{apitype},
  columns=fullflexible, keepspaces=true, breaklines=false,
  aboveskip=3pt, belowskip=1pt, xleftmargin=0pt,
  literate=%
    {<<<}{{{\color{atomicC}<{}<{}<}}}3 {>>>}{{{\color{atomicC}>{}>{}>}}}3
    {⊢}{{$\vdash$}}1 {∗}{{$\ast$}}1 {∀}{{$\forall$}}1
    {∃}{{$\exists$}}1 {↑}{{$\uparrow$}}1 {γ}{{$\gamma$}}1
    {↦}{{$\mapsto$}}1 {⌜}{{$\ulcorner$}}1 {⌝}{{$\urcorner$}}1
    {∧}{{$\land$}}1 {∨}{{$\lor$}}1 {→}{{$\rightarrow$}}1
    {¬}{{$\neg$}}1 {≠}{{$\neq$}}1 {≤}{{$\leq$}}1 {≥}{{$\geq$}}1
    {∈}{{$\in$}}1 {∉}{{$\notin$}}1 {⊆}{{$\subseteq$}}1
    {∪}{{$\cup$}}1 {∩}{{$\cap$}}1 {∅}{{$\emptyset$}}1
    {λ}{{$\lambda$}}1 {Φ}{{$\Phi$}}1 {Ψ}{{$\Psi$}}1
    {σ}{{$\sigma$}}1 {τ}{{$\tau$}}1 {α}{{$\alpha$}}1
    {β}{{$\beta$}}1 {δ}{{$\delta$}}1 {ε}{{$\epsilon$}}1
    {ρ}{{$\rho$}}1 {ω}{{$\omega$}}1 {Λ}{{$\Lambda$}}1
    {⊤}{{$\top$}}1 {⊥}{{$\bot$}}1 {≡}{{$\equiv$}}1
    {⟨}{{$\langle$}}1 {⟩}{{$\rangle$}}1 {⋅}{{$\cdot$}}1
    {▷}{{$\triangleright$}}1 {□}{{$\square$}}1 {↪}{{$\hookrightarrow$}}1
}

\definecolor{atomicC}{rgb}{0.75,0.10,0.10}   
\definecolor{apicomment}{rgb}{0.16,0.56,0.24}
\definecolor{apitype}{rgb}{0.00,0.00,0.90}
\lstdefinestyle{apibox}{%
  basicstyle=\scriptsize\ttfamily,
  commentstyle=\color{apicomment},
  keywordstyle=\color{apitype},
  morecomment=[l]{//},
  keywords={ReplyMsg,Request,LogEntry,int,bool,void},
  columns=fullflexible, keepspaces=true, breaklines=false,
  aboveskip=0pt, belowskip=0pt,
}

\makeatletter
\g@addto@macro\normalsize{%
  \abovedisplayskip=3pt plus 1pt minus 1pt
  \belowdisplayskip=3pt plus 1pt minus 1pt
  \abovedisplayshortskip=0pt plus 1pt
  \belowdisplayshortskip=2pt plus 1pt minus 1pt}
\makeatother
\makeatletter
\renewcommand\section{\@startsection{section}{1}{\z@}%
  {-2.5ex}
  {2.3ex \@plus .2ex}
  {\reset@font\large\bf}}
\makeatother

\makeatletter
\def\@maketitle{\newpage
 \vbox{%
  \vskip 1.2em
  \begin{center}%
   {\Large\bf \@title \par}%
   \vskip 0.18in
   {\large\it \lineskip .5em \begin{tabular}[t]{c}\@author\end{tabular}\par}%
  \end{center}%
  \par\vskip 0.2in}}
\makeatother

\makeatletter\renewcommand\subsection{\@startsection{subsection}{2}{\z@}{-2.25ex}{1ex \@plus .2ex}{\normalfont\large\bfseries}}\makeatother

\newcommand{\NOPaxosV}{\mbox{NOPaxos-v}\xspace}
\newcommand{\CorfuV}{\mbox{Corfu-v}\xspace}
\newcommand{\LazyLogV}{\mbox{LazyLog-v}\xspace}

\usepackage{caption}

\AtBeginDocument{%
  }

\begin{document}




\date{}

\title{\fontsize{13.5}{16}\selectfont \bf \sysname: Verifying Sequencer-Aided Coordination Systems with Cross-View Prophecy}

\author{
{\rm Jinkun Geng$^{1}$ \quad Muhammad Haseeb$^{1}$ \quad Daniel Qian$^{1}$}\\
{\rm Joseph Tassarotti$^{1}$ \quad Shuai Mu$^{2}$ \quad Anirudh Sivaraman$^{1}$}\\
$^{1}$New York University \qquad $^{2}$Stony Brook University
}

\maketitle
\begin{abstract}

Many distributed systems achieve high performance by relying on dedicated sequencers to order requests, leaving replicas to commit them in that order. Despite more than a decade of such \emph{sequencer-aided coordination (SAC)} systems, from Corfu and NOPaxos to LazyLog and SwitchBFT, their correctness has been argued only with pen-and-paper proofs or bounded model checking at the protocol level.

In this paper, we formally verify implementations of SAC protocols. The sequencer seems to make this easy: it decides the order of all requests, so the proof can place each request's linearization point at the sequencer using a well-known technique, prophecy variables. Yet a linearization point placed at the sequencer is not durable, because the sequencer orders a request before any replica commits it. Crashes and other rare events may trigger a view change that discards the request, nullifying the earlier decision and the proof that rests on it.

Our framework, \sysname{}, extends prophecy variables to handle view changes, and we make three contributions. (1) We develop \emph{cross-view prophecy peek}, which lets the proof follow a request across view changes until it is either committed or permanently discarded. (2) Our verification uncovered bugs in existing SAC implementations, including one that can lose committed requests. We fix them and prove the fixes correct. (3) We apply \sysname{} to NOPaxos, Corfu and LazyLog. We build verified implementations with machine-checked proofs of linearizability. They perform within 17\% of their open-source, unverified counterparts. The verified implementations and proofs will be open-sourced.

\end{abstract}
\pagestyle{plain}


\section{Introduction}
\label{sec:intro}

\setlength{\epigraphwidth}{0.86\columnwidth}
\epigraph{\itshape I went forward in time, to view alternate futures, to see all the possible outcomes of the coming conflict.}{Dr.\ Strange, \textit{Avengers: Infinity War} (2018)}

\noindent Many practical distributed systems, including databases~\cite{vldb20-tidb,cosmosdb,sigmod20-cockroachdb,spanner}, transaction processing systems~\cite{sigmod12-calvin,osdi16-nopaxos,sosp17-eris,nsdi23-hydra,sosp25-tiga}, shared logs~\cite{nsdi12-corfu,nsdi17-vcorfu,nsdi20-scalog,jsys23-manson, sosp24-lazylog}, and distributed storage systems~\cite{nsdi18-netchain, osdi20-pegasus,sosp21-mind, fast21-concordia,sigmod25-p4kvs}, require strong consistency in the form of linearizability. A linearizable system greatly simplifies application development by enabling developers to reason about system behavior using a simple, sequential abstraction. However, ensuring linearizability is widely acknowledged to be challenging, motivating decades of research on distributed coordination protocols. Well-studied protocols such as Paxos~\cite{paxos} and Viewstamped Replication (VR)~\cite{viewstamp} have undergone extensive refinement and rigorous testing in production, alleviating most concerns about their correctness.

However, canonical protocols such as Paxos and VR concentrate ordering, replication, and commitment at the replicas, and at the leader replica in particular, which becomes a bottleneck as load grows. This has motivated a line of work that offloads request ordering from replicas to independent entities, which we refer to as \emph{sequencers}. Sequencers take various forms, from specialized hardware such as programmable switches to general-purpose software on end hosts. Centralizing ordering at the sequencer decouples ordering from replication and removes that bottleneck, yielding substantially higher performance. The design has therefore been adopted by a wide range of systems, including Calvin~\cite{sigmod12-calvin}, Corfu~\cite{nsdi12-corfu,tocs13-corfu,nsdi17-vcorfu}, NOPaxos~\cite{osdi16-nopaxos}, LazyLog~\cite{sosp24-lazylog}, NeoBFT~\cite{sigcomm23-neobft}, SwitchBFT~\cite{nsdi26-switchbft} and many others~\cite{sosp17-eris,nsdi23-hydra,vldb23-compartmentalized-paxos}. We refer to them as \emph{sequencer-aided coordination (SAC) protocols}.

The sequencer-aided design does not eliminate the complexity of coordination but shifts it to the other cases where messages are delayed, reordered, or dropped after leaving the sequencer, or where replicas or sequencers crash. Correctness therefore does not follow automatically from the common case to cover the whole protocol. Existing correctness arguments, whether pen-and-paper proofs or bounded model checking, can miss subtle bugs. Formal verification can rule out such bugs by producing machine-checked proofs of correctness for implementations of these protocols. Indeed, during our own verification effort, we uncovered bugs in several published SAC protocols (\S\ref{sec:at-most-once-resolution}, \S\ref{sec:other-bugs}).

This paper presents \sysname{}, a general framework for formally verifying implementations of SAC protocols. To verify that these systems are linearizable, a formal proof must identify each operation's linearization point, i.e. the moment during an operation's execution at which it can be logically thought of as taking effect. Linearizability requires that the system's behavior is consistent with the ordering of operations induced by the linearization points. Since determining the order of the linearization points is a key part of a linearizability proof, a natural candidate is to choose the linearization points to be at the sequencer at the moment it decides the relative ordering of an event. The replicas, by contrast, cannot play this role: in NOPaxos, for example, a request is committed once the client collects a quorum of replies, so not even the leader knows when a request commits.

While the sequencer makes it easy in a proof to decide the relative ordering of linearization points, there is an obstacle to choosing the sequencer as the linearization point in a proof. At the sequencer, the proof cannot yet determine whether a sequenced request will eventually be committed. After leaving the sequencer, requests may be delayed, reordered, or even dropped before reaching the servers. At the same time, common proof techniques require the proof to identify each linearization point at the moment it happens, as the proof steps through the code. At the sequencer, that moment arrives before the request's fate is known: the proof must decide on the spot whether the request takes effect, yet it cannot tell whether the request will ever be committed. \emph{Prophecy variables}~\cite{tcs91-abadi-lamport}, a technique supported by several verification frameworks~\cite{popl19-prophecy-variable,sosp23-grove,vstd-lib}, address exactly this obstacle. At a high level, prophecy variables allow the proof at the sequencer(s) to look ahead at the future actions taken by the servers. Based on these prophesied outcomes, the proof can derive whether a sequenced request will eventually be committed and update the abstract state accordingly. Later, when a server actually processes the request, the proof \emph{resolves} the prophecy variable by recording the server's actual action.

However, using prophecy variables directly to verify SAC systems is not straightforward, particularly in the presence of crashes and recovery. The central technical challenge arises from the decoupling between \emph{ordering} and \emph{durability}: one or more sequencers first assign a tentative order to requests, after which servers try to commit the requests according to that order, so a linearization point placed at the sequencer is not durable. Like VR, SAC protocols recover from crashes through a \emph{view change}: the system moves to a new view and rebuilds its state from the replicas that take part in recovery. Crashes introduce additional uncertainty: \emph{if a request is sequenced but not yet committed when a crash occurs, will it eventually be committed after recovery, or will it be abandoned?} An uncommitted request may either (1) survive and later be committed, or (2) be permanently discarded during recovery, and which outcome occurs depends on which replicas take part in the view change. As a result, the proof at the sequencer must make its decision in the presence of uncertainty about both future commitment and recovery behavior.

\Para{Cross-view prophecy peek.}
To address this, \sysname{} develops \emph{cross-view prophecy peek}, which adopts a view-based abstraction built on top of prophecy variables that is inspired by VR~\cite{viewstamp}. Cross-view prophecy peek records a separate prophesied value for each view, which confines the non-determinism of crash recovery to one view at a time. Consider an uncommitted request whose fate is uncertain in the current view. Cross-view prophecy peek allows the proof to look ahead at the request's assigned slot across subsequent views until it finds a view in which the request is either committed by a majority of servers or permanently replaced by a no-op. This enables the proof at the sequencer to soundly determine the outcome for the linearizability argument.

\Para{Identifying bugs and escrow technique.}
Prophecy variables also impose an additional discipline on the proof. In several verification frameworks, prophecy variables have an associated notion of permission or ownership, which is required to \emph{resolve} a prophecy variable, i.e. to attach the prophesied value to the program event it is designed to make predictions about. Just as Rust's ownership system rules out certain forms of races, this ownership discipline for prophecies is needed to make it clear \emph{which} event is being predicted~\cite{popl19-prophecy-variable}. While verifying several SAC implementations with our technique, we realized that this ownership discipline was being violated in our initial proof attempts, which uncovered bugs in the actual implementations. For example, in NOPaxos, a replica that crashes may rejoin the \emph{same} view after a state transfer and reprocess slots it had already processed. In our proof, this would have required changing which value the prophecy's prediction was about, which these ownership rules forbid. This is not merely a proof technicality. It corresponds to a concrete bug in which the client is told that a request is committed, yet the request is lost during recovery. Similar issues exist in other SAC protocols (e.g., Eris~\cite{sosp17-eris}, Hydra~\cite{nsdi23-hydra}). To fix it, we adopt the \emph{escrow} technique~\cite{oopsla14-gps-escrow,grove-thesis}: when restarting, each replica draws a \emph{fresh, non-duplicable token} at boot, and the coordinator releases a view's resolution permission to the replica in exchange for the one token. A replica that crashes and reboots holds a different token, so it cannot reclaim that permission to resolve prophesied values in that view.

\Para{Our contributions.}
We implement \sysname in Grove~\cite{sosp23-grove} and demonstrate it on three representative SAC systems: NOPaxos, Corfu and LazyLog. \sysname makes three contributions:

(1) \sysname{} introduces \emph{cross-view prophecy peek}, a novel verification technique that is generally applicable to SAC systems. It resolves the fundamental uncertainty arising from the decoupling of ordering and durability under crash recovery.

(2) \sysname{} exposes multiple bugs in existing SAC protocol implementations. The most consequential one surfaced through the prophecy discipline: a replica that rejoins its previous view resolves a prophecy twice and loses a committed request.  We develop our remedy based on the escrow technique and prove the fix is correct.

(3) \sysname{} delivers machine-checked proofs and verified implementations of NOPaxos, Corfu and LazyLog. Our evaluation shows that the verified implementations perform comparably to their open-source, unverified counterparts (within 17\%).

\section{Overview of SAC Protocols}
\label{sec:protcol}

In this section, we provide background on SAC protocols by describing their common design pattern and highlighting where individual protocols diverge. 

In traditional coordination protocols like Paxos and VR, the leader replica is responsible for ordering, replicating, and committing requests, leading to a performance bottleneck. SAC protocols avoid this by decoupling ordering from replication: one or more dedicated sequencers establish a tentative order for incoming requests ahead of replication, while replicas focus solely on processing and committing requests in that order. In the common case, replicas simply follow the sequencer-determined order, making the sequencer(s) the effective linearization point.

However, the tentative order may not always hold. Requests can be delayed, reordered, or dropped after leaving the sequencer, and replicas or sequencers may crash. SAC protocols therefore incorporate additional mechanisms to handle these abnormal cases while preserving linearizability.

\begin{figure*}[!t]
    \centering
	\begin{subfigure}[t]{0.32\textwidth}
		\centering
		\includegraphics[width=\linewidth]{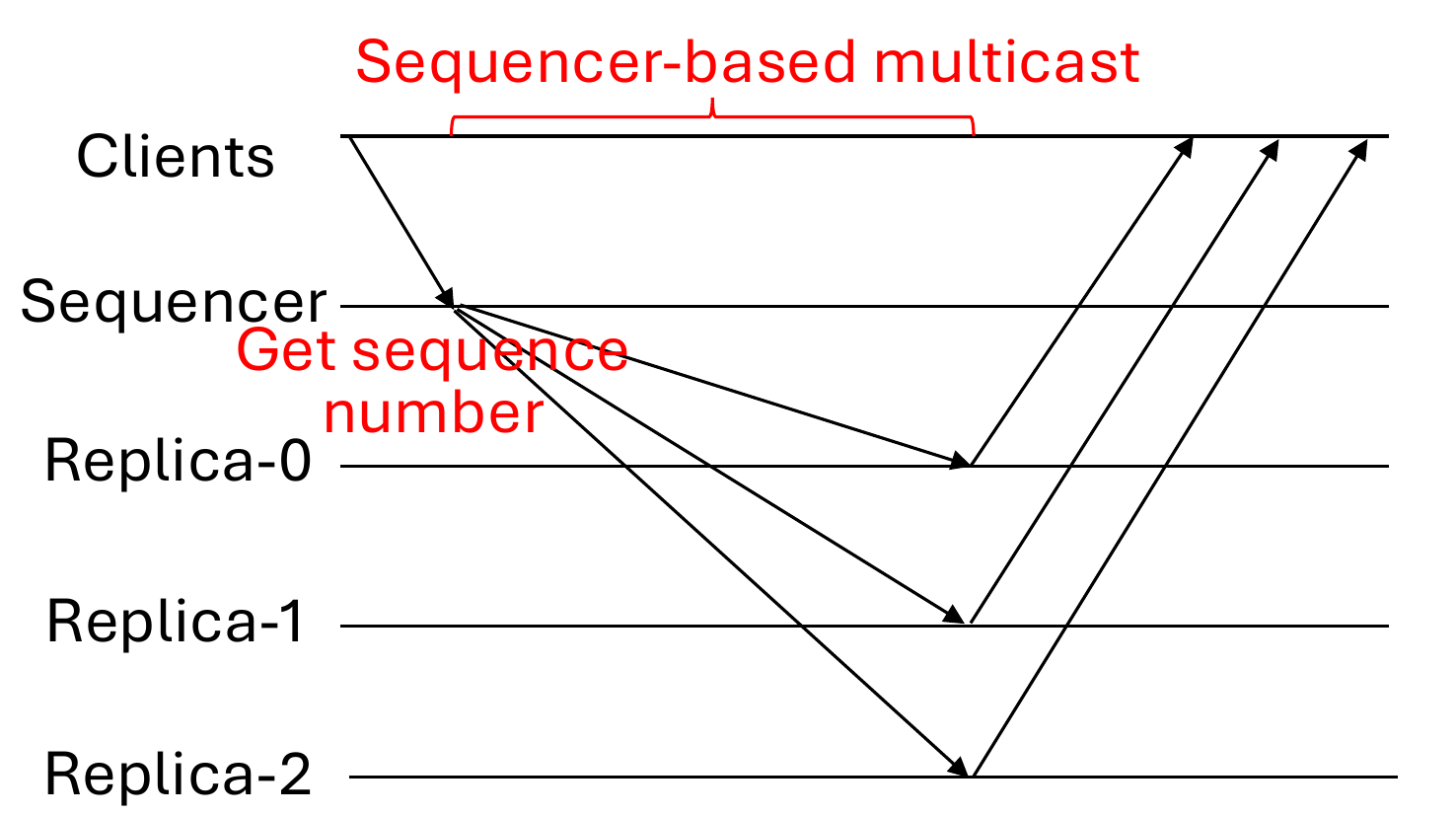}
		\caption{NOPaxos}
		\label{fig:nopaxos-workflow}
	\end{subfigure}
    \hfill
    \begin{subfigure}[t]{0.32\textwidth}
		\centering
		\includegraphics[width=\linewidth]{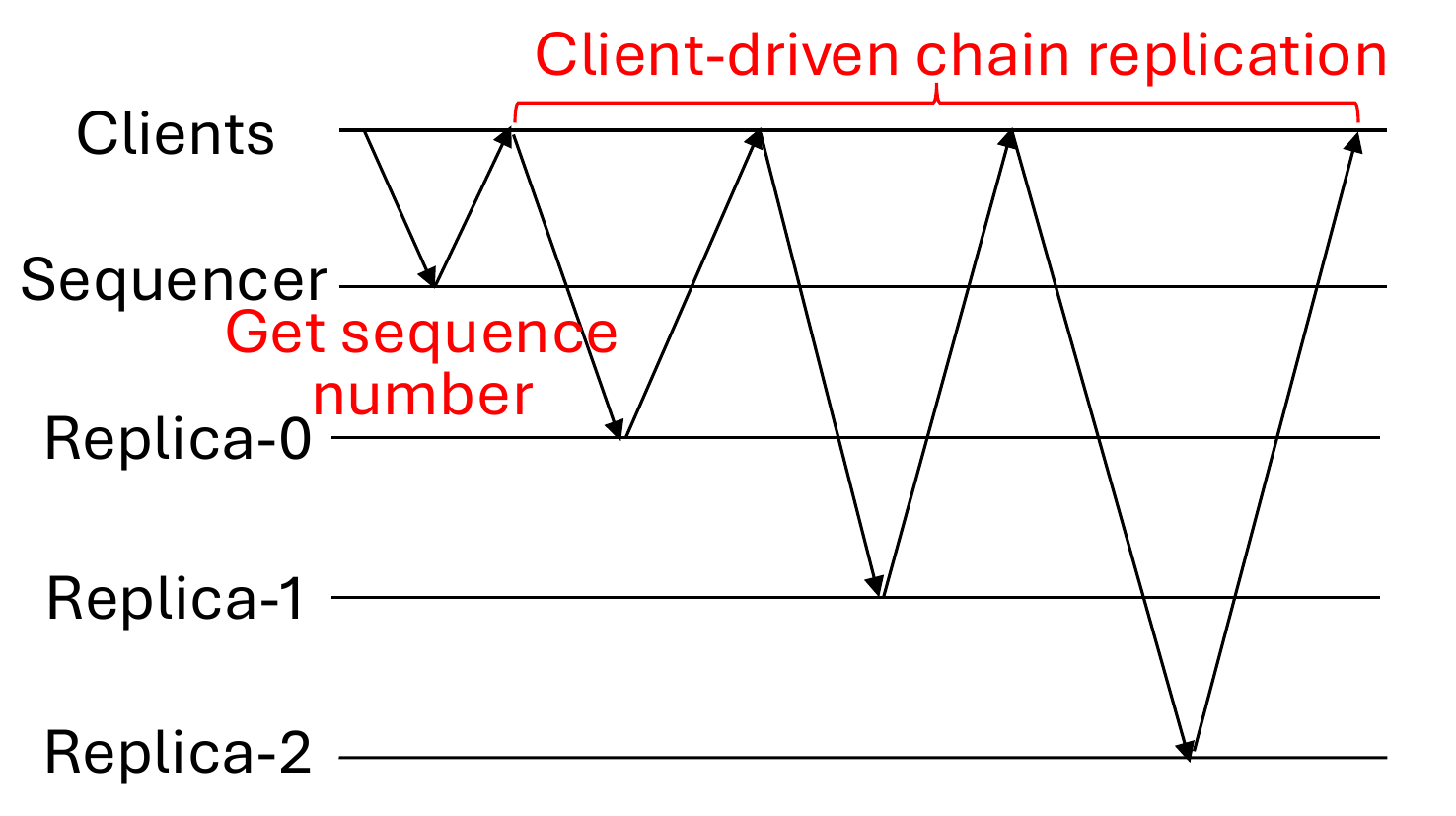}
		\caption{Corfu}
		\label{fig:corfu-workflow}
   \end{subfigure}
    \hfill
    \begin{subfigure}[t]{0.32\textwidth}
        \centering
        \includegraphics[width=\linewidth]{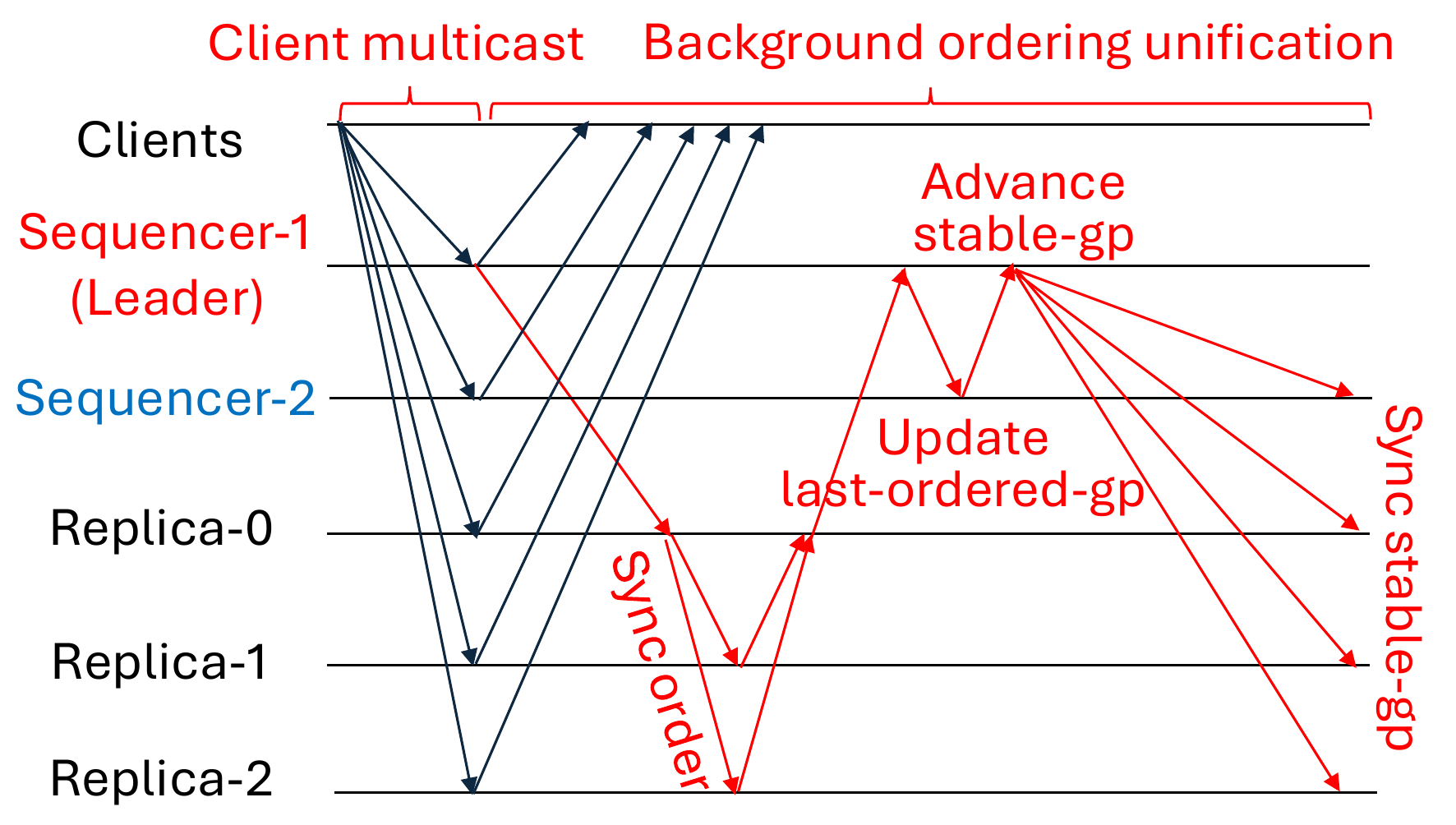}
        \caption{LazyLog}
        \label{fig:lazylog-workflow}
    \end{subfigure}
    \caption{Workflows of three representative SAC protocols. 
    }
    \Description{Three side-by-side workflow diagrams showing how a request flows through the sequencer, replicas, and client in NOPaxos, Corfu, and LazyLog.}
    \label{fig:sac-workflows}
\end{figure*}

We illustrate this common structure and key design differences using three typical protocols:  NOPaxos~\cite{osdi16-nopaxos}, Corfu~\cite{nsdi12-corfu}, and LazyLog~\cite{sosp24-lazylog}. Figure~\ref{fig:sac-workflows} shows their workflows.

\Para{Sequencing and ordering.}
In all three protocols, the sequencer assigns each request a monotonically increasing sequence number that determines the tentative ordering. The protocols differ in how requests reach replicas after sequencing. In NOPaxos, the sequencer directly multicasts each request to all replicas. In Corfu, the sequencer returns the sequence number to the client, which then delivers the request to replicas in a client-driven chain-replication style. LazyLog employs multiple sequencers; clients multicast request metadata to all sequencers and send request data directly to all replicas. Among the sequencers, one is designated as the leader and determines the linearizable order. (We refer to LazyLog's Erwin-st variant~\cite{sosp24-lazylog} with a single shard.)

\Para{Commitment.}
In the common case, replicas process requests in sequence-number order and send replies to clients. NOPaxos commits once the client collects replies from a quorum of replicas. Corfu commits a request once all replicas in the chain acknowledge it. LazyLog commits once the client receives replies from all sequencers and all replicas.

\Para{Inconsistency handling.}
When requests are delayed, reordered, or dropped after sequencing, replicas may not observe the tentative order, and each protocol resolves the resulting inconsistencies differently. In NOPaxos, the leader replica runs \emph{gap agreement}, contacting a quorum to mark the missing slot \texttt{NOOP} so that replicas can move on. In Corfu, later requests fill the missing slots with junk values, which can freeze into \texttt{NOOP}s by chain replication, so that a late request can no longer occupy the slot even if it eventually arrives. In LazyLog, the leader sequencer reconciles the order in the background: once replicas agree, it advances a local counter (\emph{last-ordered-gp}), then synchronizes across sequencers and advances a second one (\emph{stable-gp}; see \S 4.3 of~\cite{sosp24-lazylog}).

\Para{Crash recovery.}
When replicas or sequencers fail, each protocol uses a different recovery mechanism. NOPaxos uses a VR-style view change to elect a new leader and recover committed requests. During recovery, uncommitted requests may either be committed or abandoned, depending on which replicas participate.  Corfu employs a reconfiguration protocol based on Vertical Paxos~
\cite{podc09-vertical-paxos} to recover committed requests and replace failed components. LazyLog elects a new leader sequencer and initiates a view-based recovery procedure. The new leader flushes its log to the replicas, potentially overwriting the order established by the previous leader while preserving linearizability.

This non-determinism during crash recovery significantly complicates correctness reasoning. Uncommitted requests may survive or be permanently discarded depending on the recovery outcome, yet the proof must account for all possibilities at the time of sequencing. We discuss how \sysname{} addresses this challenge in \S\ref{sec:prove-protocol}.

\section{Verification with Prophecy Variables}
\label{sec:prove-protocol}

This section provides an overview for \emph{prophecy variables}~\cite{popl19-prophecy-variable}, and explains how prophecy variables can support the proof that depends on future events. However, we will continue to show, simply adopting the conventional prophecy variable design is insufficient to handle the non-determinism when verifying  SAC systems with crash recovery, which motivates our design of cross-view prophecy peek (\S\ref{sec:cross-view}).

\subsection{Overview of Prophecy Variables}

Many program verification techniques proceed line-by-line through programs, reasoning about how the system's state would evolve as each statement of the program is executed.
A prophecy variable allows us to know, during the proof, what the eventual value of a program expression will be before getting to the place where that expression executes.

As a simple example adapted from~\cite{popl19-prophecy-variable}, consider a \emph{lazy coin}. Figure~\ref{fig:lazy-coin} shows two implementations. An eager coin (left) flips immediately at creation time and stores the result. A lazy coin (right) defers the flip: \texttt{new\_coin()} creates the coin with value \texttt{None}, and the actual flip only happens on the first call to \texttt{read\_coin()}. From a client's perspective, the two are indistinguishable, so they should have the same specification: after creation, there exists some boolean $b$ such that every read returns $b$.

\begin{figure}[!t]
\centering
\hrule
\vspace{4pt}
\begin{minipage}[t]{0.46\columnwidth}
\begin{algorithmic}[1]
\Statex \textbf{Eager coin}
\Function{new\_coin}{}
    \State $c.val \gets \text{toss}()$
    \State \Return $c$
\EndFunction
\Statex $\{c.\ \exists b.\ \textsf{Coin}(c, b)\}$
\vspace{4pt}
\Function{read\_coin}{$c$}
    \State \Return $c.val$
\EndFunction
\Statex $\{v.\ v = b \ast \textsf{Coin}(c, b)\}$
\end{algorithmic}
\end{minipage}
\hfill
\begin{minipage}[t]{0.50\columnwidth}
\begin{algorithmic}[1]
\Statex \textbf{Lazy coin}
\Function{new\_coin}{}
    \State $c.val \gets \text{None}$
    \State \textcolor{gray}{$c.p \gets \textbf{NewProph}()$}
    \State \Return $c$
\EndFunction
\Statex $\{c.\ \exists b.\ \textsf{Coin}(c, b)\}$
\vspace{4pt}
\Function{read\_coin}{$c$}
    \If{$c.val = \text{None}$}
        \State $r \gets \text{toss}()$
        \State $c.val \gets r$
        \State \textcolor{gray}{$\textbf{Resolve}\ c.p\ \textbf{to}\ r$}
    \EndIf
    \State \Return $c.val$
\EndFunction
\Statex $\{v.\ v = b \ast \textsf{Coin}(c, b)\}$
\end{algorithmic}
\end{minipage}
\vspace{4pt}
\hrule
\vspace{4pt}
\hspace{-5em}$\textsf{Coin}(c, b) \triangleq (c.val = \text{Some}\ b)$\\\hspace{4em} $\;\vee\; (c.val = \text{None} \ast \textsf{Proph}(c.p, b))$
\caption{Eager and lazy coin implementations (adapted from~\cite{popl19-prophecy-variable}). Gray lines are ghost code that exists only in the proof. \textbf{NewProph} creates a prophecy variable with an existentially quantified value. \textbf{Resolve} records the actual outcome. The predicate $\textsf{Coin}(c, b)$ connects the prophecy variable to the postcondition.}
\Description{Pseudocode of the eager and lazy coin implementations, with the ghost prophecy operations shown in gray.}
\label{fig:lazy-coin}
\end{figure}

The verification challenge arises when proving \texttt{new\_coin()} for the lazy implementation. The postcondition must state that the coin has some value $b$, but $b$ will not be determined until the first \texttt{read\_coin()} call. The proof cannot instantiate $b$ at creation time, so it gets stuck. A prophecy variable resolves this (line 3, right): at \texttt{new\_coin()}, the proof introduces a prophecy variable to represent the future flip result. The postcondition can now be stated in terms of this prophesied value. Later, when \texttt{read\_coin()} performs the actual flip (line 10, right), the prophecy is \emph{resolved}: the prophesied value must equal the concrete flip result in order for the proof to complete.



These prophecy variables are just \emph{ghost code} annotations that are used for tracking information in the proof.
They are erased during compilation and have no effect on the program's actual behavior at run time.

\subsection{Prophecy Variables for SAC Protocols}

To prove linearizability for SAC protocols, \sysname{} maintains a \emph{ghost log} that records the abstract sequential history of committed requests. Each time the sequencer assigns a sequence number to a request, the proof must decide whether to append the request to this ghost log. If the request will eventually be committed by the replicas, it should be appended; if the request will ultimately be uncommitted (e.g., dropped), it should not. This decision must be made when the sequencer assigns the sequence number, because the sequencer is the linearization point where the request's position in the ghost log is determined.

The difficulty is that when the sequencer assigns the sequence number, the request has not yet reached the replicas. The request may arrive in order and form a quorum, or it may be delayed, reordered, or dropped. Replica crashes add further uncertainty. Without prophecy variables, the proof at the sequencer step has no direct way to refer to the replicas' future actions, and thus cannot determine whether to extend the ghost log.

To address this, for each replica and each request slot, \sysname{} introduces a prophecy variable for the replica's future action for that request: either \texttt{Normal} (place the request at this slot) or \texttt{NOOP} (mark the slot as a no-op operation). This lets the proof reason in two stages. At the sequencer step, the proof examines (\emph{peeks} at) the prophecy variables across replicas. Since these variables reflect the replicas' eventual actions in the current execution, the proof can determine whether the request will be committed and decide accordingly whether to extend the ghost log. At the replica step, the proof \emph{resolves} the prophecy variable by recording the replica's actual action. Since the prophesied value reflects what actually happens in each execution, the two always agree.

For example, consider NOPaxos. After the sequencer multicasts a request to replicas, the request is committed only if it arrives at a majority of replicas (including the leader) in the sequencer-assigned order. However, a subsequent request with a larger sequence number may arrive at replicas earlier, while the current request is delayed or dropped. In such a scenario, the NOPaxos leader initiates \emph{gap agreement}, contacting a majority of replicas to mark the missing slot as \texttt{NOOP}. Once complete, the original request can no longer be committed at that slot. These two outcomes (commit vs.\ no-op) correspond to different execution traces, each with different prophesied values in the replicas' prophecy variables. Similar non-determinism arises in other SAC protocols such as Corfu, where later requests fill gaps with junk values.

\subsection{Derivation with Prophecy Variables}

\begin{figure}[!t]
    \centering
    \includegraphics[width=0.8\linewidth]{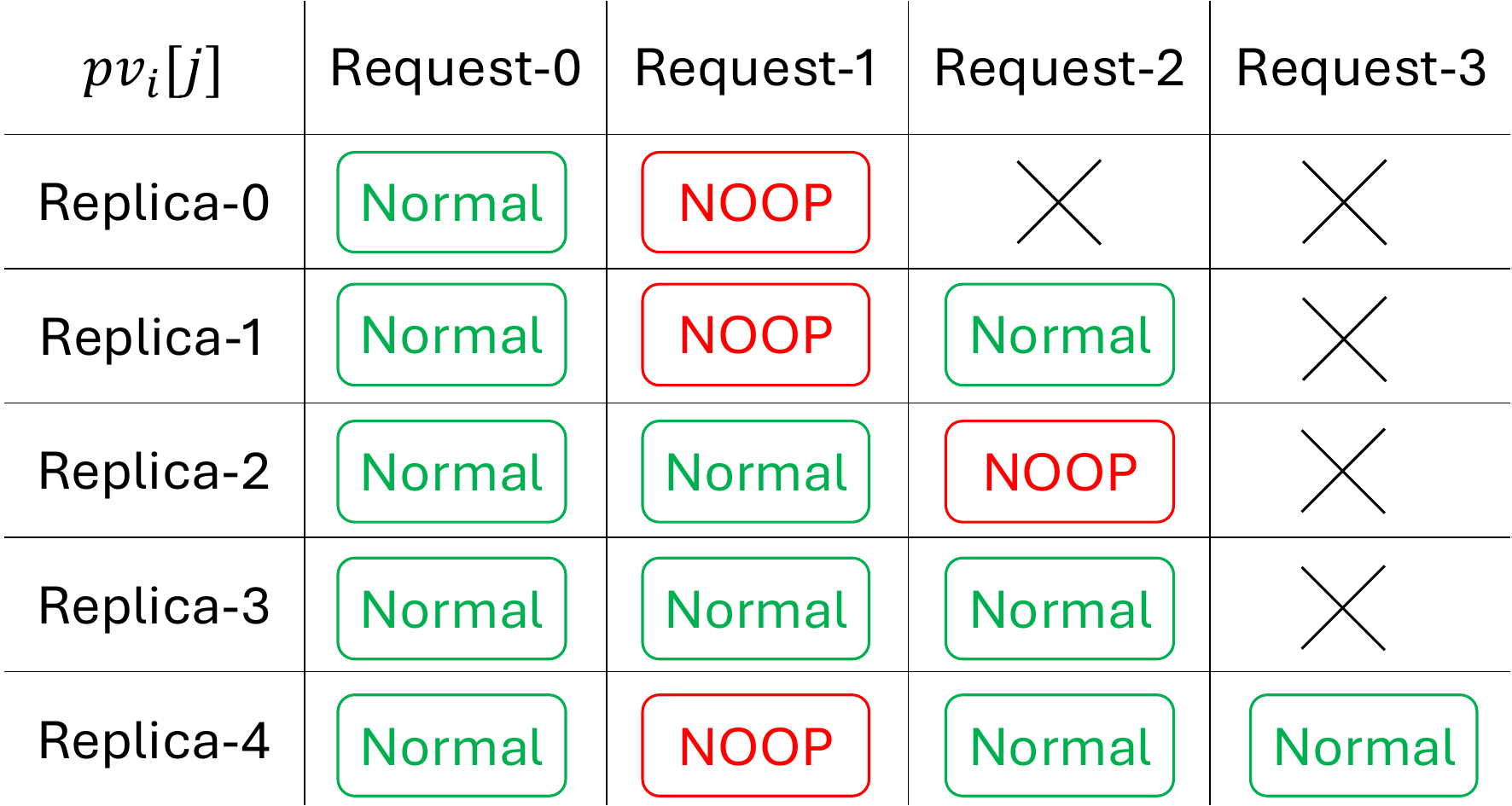}
    \caption{Prophecy variable design for NOPaxos. Each row represents the prophecy list of one replica, which contains the prophesied replica action for each request. Replica-0 is the leader replica. This single-view design cannot resolve ambiguity introduced by crash recovery (see Request-2).}
    \Description{Table of prophesied actions per replica and request for NOPaxos, illustrating a committed request, a no-op request, and a request left ambiguous by the leader's crash.}
    \label{fig:nopaxos-prophecy-example}
\end{figure}
We now illustrate the prophecy variable design in more detail using NOPaxos as an example. Every Replica-$r$ owns a \emph{sequence prophecy variable}~\cite{popl19-prophecy-variable}, i.e., a list of prophesied values (pvs) that describe the replica's future behavior.  Let $pv_r$ denote Replica-$r$'s prophecy list; the $j$th element $pv_r[j]$ is either \texttt{Normal} (request $j$ arrives and is appended) or \texttt{NOOP} (request $j$ is dropped and the slot is marked via gap agreement). The sequencer peeks at these values across all replicas to derive each request's fate. Figure~\ref{fig:nopaxos-prophecy-example} illustrates this with $2f+1=5$ replicas ($f=2$), where Replica-0 is the leader. We walk through three requests.

\Para{Request-0 (committed).} The leader (Replica-0) and at least $f=2$ followers (Replica-1, Replica-2) have $pv_r[0] = \texttt{Normal}$. Since a majority including the leader will append the request, the proof concludes that Request-0 will be committed and extends the ghost log with a \texttt{Normal} entry.

\Para{Request-1 (no-op).} The leader's prophecy list indicates $pv_0[1] = \texttt{NOOP}$. This corresponds to an execution where the leader does not receive Request-1 and completes gap agreement for that slot. Regardless of what the followers do (i.e., either they receive Request-1 or not), the slot for Request-1 becomes a no-op. Correspondingly, the proof extends ghost log with a \texttt{NOOP} entry instead of a \texttt{Normal} entry.


\Para{Request-2 (ambiguous).} Replica-0's prophecy list has no entry for Request-2, because Replica-0 crashes before resolving it. Request-2 arrives at Replica-1, Replica-3, and Replica-4 (all record \texttt{Normal}), but is dropped at Replica-0 and Replica-2. Before crashing, Replica-0 initiates gap agreement; Replica-2 receives the leader's gap instruction and marks the slot as \texttt{NOOP}. At this point, the prophecy lists do not contain enough information to determine Request-2's fate: depending on which replicas participate in crash recovery, the request may be either committed or discarded. This ambiguity cannot be resolved with the current single-view prophecy design, motivating the cross-view extension in \S\ref{sec:cross-view-prophecy}.

\section{Cross-View Prophecy Peek}
\label{sec:cross-view}

The previous section showed how prophecy variables enable the proof to derive request outcomes, but also revealed that the single-view design cannot resolve ambiguity when replicas crash (Request-2 in \S\ref{sec:prove-protocol}). Crash recovery introduces further non-determinism: the same request may be committed or discarded depending on which replicas participate in recovery. This section presents \emph{cross-view prophecy peek}, our key technique for reasoning about SAC protocols under crash recovery. We also describe the bugs we discovered in existing protocols and the escrow-based fix.

\subsection{Non-Determinism in Crash Recovery}
\label{sec:non-determinism-crash}

We continue with the NOPaxos example from \S\ref{sec:prove-protocol}. In Figure~\ref{fig:nopaxos-prophecy-example}, after the leader Replica-0 crashes, the remaining replicas elect a new leader, which contacts $f$ replicas (to form a $f+1$ majority) to recover the log. The recovery procedure guarantees that all previously committed requests (e.g., Request-0) are recovered and that all slots that were previously agreed as gaps remain gaps (e.g., Request-1). For requests that were neither successfully committed nor conclusively marked as gaps before the crash (e.g., Request-2), the recovery outcome is non-deterministic; it depends on which replicas participate in the recovery and what messages they receive first.

Consider Request-2 as an example. Suppose Replica-1 is elected as the new leader. If Replica-1 contacts Replica-3 and Replica-4 during recovery, then all replicas in the recovery quorum record Request-2 as a normal entry, and Request-2 is eventually committed. In contrast, if Replica-1 contacts Replica-2 and Replica-3, then Replica-1 observes a \texttt{NOOP} entry in Replica-2's log. 
Following NOPaxos's recovery rules (\S 5.2.3 in the NOPaxos paper~\cite{osdi16-nopaxos}), Replica-1 will mark Request-2 to be \texttt{NOOP} in the recovered log. Thus, the same request may either be committed or discarded depending solely on the recovery quorum.

Similar non-determinism arises for subsequent requests. For example, Request-3 is multicast by the sequencer but arrives only at Replica-4 and is dropped by the other replicas. If Replica-4 participates in recovery, then Request-3 may be recovered and eventually committed. If Replica-4 does not participate, then Request-3 does not appear in the recovered log at all. In such executions, after the sequencer multicasts Request-4, it should not make any update to the ghost log for Request-3, since the request is never committed.

These examples illustrate that crash recovery introduces non-determinism that the single-view prophecy design from \S\ref{sec:prove-protocol} cannot handle. The challenge is not simply that the future is unknown, as in the case without recovery. Rather, crash recovery makes the future \emph{non-deterministic}: for the same request, different recovery quorums yield different outcomes. A single prophesied value per replica per request cannot capture this, because replicas may take different actions across different views. To address this, we extend the prophecy variable design to track replica actions \emph{per view}, enabling the proof to follow a request's fate across view changes.

\subsection{Cross-View Prophecy Peek Design}
\label{sec:cross-view-prophecy}

\begin{figure}[!t]
    \centering
    \includegraphics[width=0.85\linewidth]{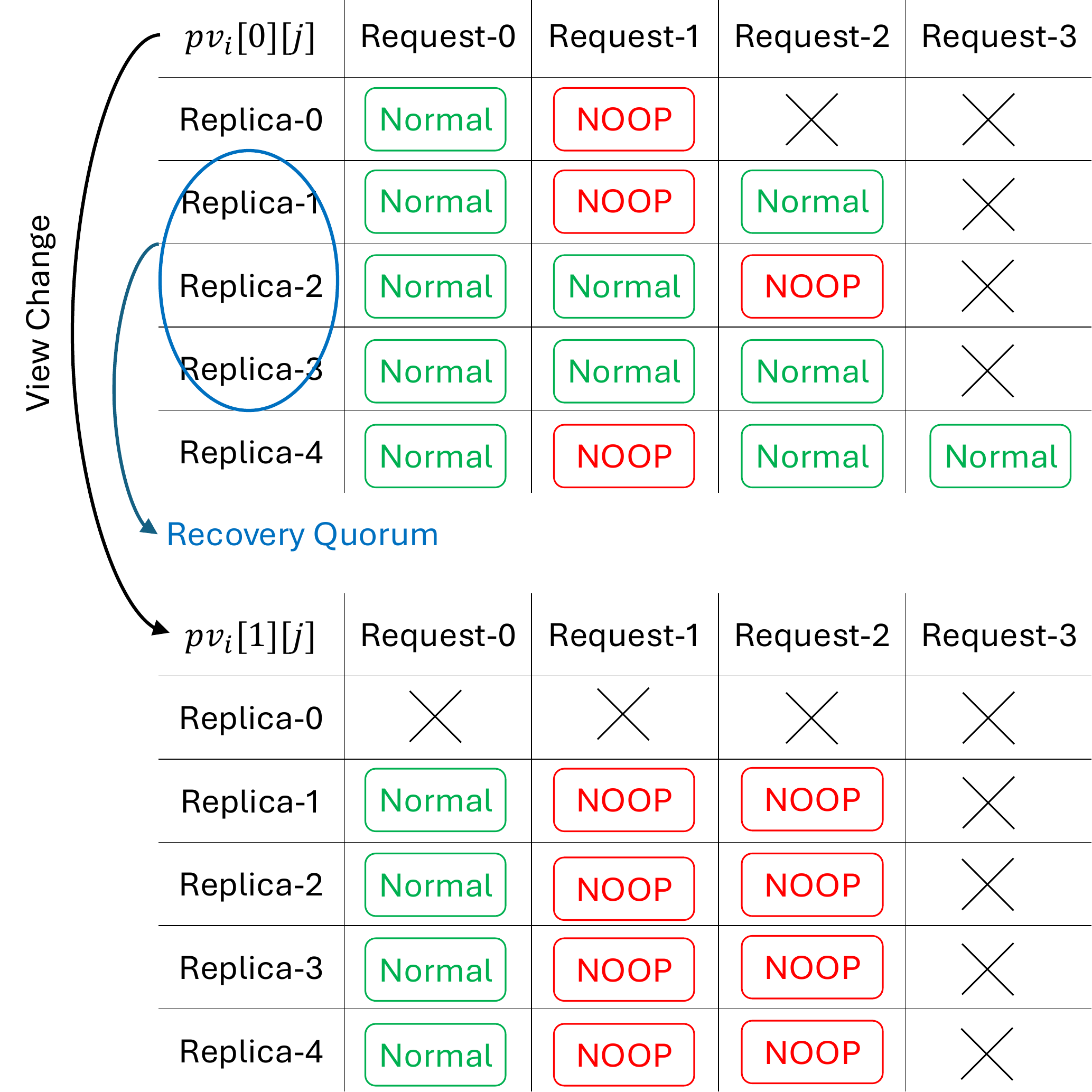}
    \caption{Cross-view prophecy variable design for NOPaxos. For simplicity, here we only consider view change and ignore session change (i.e., triggered by sequencer's crash recovery). Replica-0 is the leader in the old view $v=0$, after it crashes, Replica-1 becomes the new leader. Replica-1, Replica-2 and Replica-3 establish the recovery quorum and recover the system state, so the recovered state for the four requests is reflected in the prophecy variables in higher view $v=1$. }
    \Description{Prophesied actions per replica for four requests across two views, showing how recovery in view 1 settles the requests left ambiguous in view 0.}
    \label{fig:nopaxos-cross-view-prophecy-example}
\end{figure}

Our design is inspired by the view-based mechanism in Viewstamped Replication~\cite{viewstamp-original}, where views provide a clear boundary for handling crash recovery. After each recovery, replicas with restored state (i.e., logs) transition into a new view.

For each request, \sysname{} does not model replica behavior using a single prophesied value. Instead, it adopts a finer-grained, view-aware design. Concretely, for each replica, \sysname{} constructs a two-dimensional prophecy list, where $pv_r[v][j]$ denotes the Replica-$r$'s action for Request-$j$ in view $v$. Since these prophecy lists reside in the global invariant and are accessible to the sequencer, the sequencer can peek at multiple prophesied outcomes for each request, enabling it to infer the eventual outcome of multicast requests. Figure~\ref{fig:nopaxos-cross-view-prophecy-example} illustrates the same four requests as in Figure~\ref{fig:nopaxos-prophecy-example}, now extended across two views ($v=0$ and $v=1$).

For Request-0 and Request-1, the sequencer can determine their eventual outcomes by inspecting the prophesied values in view $v=0$. However, for Request-2 and Request-3, Replica-0 (the leader in $v=0$) no longer provides sufficient information, preventing the sequencer from conclusively determining their outcomes. Unlike the setting in Figure~\ref{fig:nopaxos-prophecy-example}, the finer-grained design allows the sequencer to continue peeking into the next view ($v=1$), where Replica-1 is the leader and forms a recovery quorum with Replica-2 and Replica-3. Based on the prophesied values in $v=1$, Request-2 is recovered as a \texttt{NOOP}, while Request-3 is discarded (since Replica-4 does not participate in recovery). Therefore, the sequencer can safely update the ghost log: it appends a \texttt{NOOP} entry for Request-2 and leaves the log unchanged for Request-3.

Cross-view prophecy peek generalizes to multiple consecutive crashes. If Replica-1 also crashes, the sequencer continues peeking into future views ($v=2, 3, \ldots$) until it can determine each request's outcome. The approach also handles sequencer crash recovery. In NOPaxos, the view identifier is a pair $\langle\mathit{sessionId}, \mathit{viewId}\rangle$, where $\mathit{sessionId}$ is incremented on sequencer recovery. Cross-view prophecy peek operates over both dimensions, allowing the proof to reason about requests across sequencer and replica failures (see \S\ref{sec:instantiate-proof}).

\subsection{\emph{At-Most-Once} Prophecy Resolution }
\label{sec:at-most-once-resolution}
While the sequencer peeks at prophecy variables to derive request outcomes, each prophesied value (pv) must eventually be \emph{resolved} when the replica processes the request.

Upon receiving a request, the replica either appends it as a \texttt{Normal} entry or marks its slot as \texttt{NOOP}. In the proof, this step resolves the corresponding prophesied value in $pv_r$. A prophecy variable is an \emph{exclusive} resource in separation logic: only its owner may resolve it, and each resolution consumes the prophesied value at the head of $pv_r$~\cite{popl19-prophecy-variable}. Consequently, each prophesied value can be resolved at most once, and only by the replica that owns $pv_r$.

Based on this \emph{at-most-once} requirement of prophecy resolution, we revisit the crash recovery procedures in open-source SAC implementations.
Among the systems we examined, few implement crash recovery at all, which is anecdotal evidence that recovery is a difficult and error-prone part of the design. For those that do include recovery (e.g., NOPaxos~\cite{nopaxos-implementation}\footnote{The NOPaxos paper~\cite{osdi16-nopaxos} claims that the recovery procedure is a direct equivalent of VR and includes the VR's recovery implementation in their open-source codebase~\cite{nopaxos-implementation}.}), we find that they do not guarantee \emph{at-most-once} resolution for every prophesied value. Specifically, in NOPaxos, a failed replica may rejoin the system in the same \emph{view} as before the crash. Because NOPaxos is an in-memory protocol without durable storage, the rejoined replica does not retain its prior behavior, even if it has already resolved prophecies in that view. As a result, the proof would have to resolve the same prophesied value a second time. The exclusivity of $pv_r$ forbids this, so the linearizability proof cannot be completed.

A failed proof attempt does not by itself imply a bug in the protocol. However, in this case,  we observe that there is an underlying concrete correctness bug in NOPaxos, as we show below with an explicit buggy trace.



\Para{Error trace in physical execution.} We consider Replica-0 as the leader in view $v=0$. 
Before crashing, Replica-$r$ ($r \neq 0$) receives all requests, appends them as normal entries, and sends replies to clients. Suppose the last request, Request-$j$, arrives at Replica-$r$ first. The remaining replicas have not yet received Request-$j$ when Replica-$r$ crashes and initiates recovery. During recovery, Replica-$r$ contacts other replicas, but since none of them have received Request-$j$, it only recovers Request-0 through Request-$(j-1)$.

We then let each follower replica follow the same crash-recovery pattern, such that each sends a reply for Request-$j$ before crashing, but fails to recover it afterward. Eventually, Request-$j$ reaches Replica-0 (the leader), which also replies in view $v=0$ before crashing. At this point, the client has collected sufficient replies to satisfy the quorum condition and thus considers Request-$j$ committed. However, after the leader crashes, none of the remaining replicas retain Request-$j$. As a result, although the client observes Request-$j$ as committed, the request is permanently lost.

\subsection{Escrow Technique}
\label{sec:escrow}

To prevent duplicate resolution, we adopt the \emph{escrow} technique~\cite{oopsla14-gps-escrow,grove-thesis}, which enables exclusive resources to be conveyed through duplicable assertions while ensuring that they can be extracted at most once.

The root cause of the bug is the lack of permission control over prophecy resolution within each view. A crashed replica can rejoin and re-resolve prophecies it already resolved. A natural fix is to require replicas to obtain a one-time permission before resolving. In practice, SAC systems already rely on a fault-tolerant coordinator (e.g., ZooKeeper~\cite{zookeeper}) to manage control-plane metadata, so we use this coordinator to manage resolution permissions as well. 
We assume the coordinator is correct and place it outside the scope of our verification.
Verified implementations of such coordinators are also available, e.g., Grove~\cite{sosp23-grove} provides a verified Paxos-based coordinator. 

The escrow pattern enforces one-shot permission transfer using two components: a duplicable invariant and an exclusive token. When Replica-$r$ starts or reboots, it creates a globally unique identifier $u$, which grants it an exclusive token $\mathit{Tok}(u)$ in the proof. The coordinator creates an escrow invariant that links the resolution permission to this token:
\[
E_{\mathit{inv}} = \boxed{\mathit{permission}(r, v) \;\vee\; \mathit{Tok}(u)}
\]
To obtain $\mathit{permission}(r, v)$, Replica-$r$ surrenders its token $\mathit{Tok}(u)$ to the escrow invariant $E_{\mathit{inv}}$ in exchange for the permission. Since $\mathit{Tok}(u)$ is exclusive (i.e., $\mathit{Tok}(u) \ast \mathit{Tok}(u) \Rightarrow \mathrm{False}$), this exchange can happen at most once.

If Replica-$r$ crashes, it loses $\mathit{permission}(r, v)$. Upon reboot, although the coordinator may re-transfer the same invariant $E_{\mathit{inv}}$, Replica-$r$ cannot regenerate the same identifier and thus does not possess the corresponding token $\mathit{Tok}(u)$. Consequently, it cannot reclaim $\mathit{permission}(r, v)$ and is prevented from resolving prophecies again in the same view $v$.

On the coordinator side, correctness requires that each $\mathit{permission}(r, v)$ is linked to at most one token. The coordinator enforces this by recording issued identifiers and allocating fresh view identifiers for new permissions. When it receives a fresh identifier, it creates a new escrow invariant; otherwise, it re-transfers the existing one.

\section{Instantiating the Proof}
\label{sec:instantiate-proof}

This section describes the concrete instantiation of \sysname{} following the order of the abstract state, prophecy resolution and prophecy peek. We take NOPaxos, Corfu and LazyLog as examples to illustrate how we instantiate proof. However, \sysname is generally applicable to other SAC systems.

\subsection{Abstract State}

Each system represents its abstract state as a \emph{ghost log}: an ordered list of entries that lives in the global invariant and is updated atomically by the sequencer alone. Every Replica-$r$ owns a sequence prophecy variable $pv_r$, which is a list of prophesied values (pvs) that describe the replica's future behavior. Because the prophecy variables also live in the global invariant, the sequencer can peek at all of them, across views. Each time a request passes through the sequencer, the sequencer-side proof peeks at the prophecy variables and performs one atomic update to the ghost log: it decides whether to append a \texttt{Normal} entry, append a \texttt{NOOP} entry, or leave the ghost log unchanged.


\subsection{Prophecy Variable Design}
\label{sec:pv-design}

In general, there are two aspects of a replica's behavior that matter for the proof: the boundaries of its lifecycle, which alternates between view changes and normal processing, and the actions it takes on individual requests during normal processing. Each sequence prophecy variable therefore contains two categories of pvs:

	(1) \textbf{Marker pv} $m = \{ v, \mathit{event}, \mathit{basePos}, \dots \}$, which marks a lifecycle boundary of the replica across views.

	(2) \textbf{Entry pv} $e = \{ v, \mathit{id}, \mathit{pos}, \mathit{action}, \mathit{rids}, \dots \}$, which records the action taken in view $v$ on the request (identified by $\mathit{id}$)  placed at log position $\mathit{pos}$.

We now explain the fields common to all protocols. Notably, a resolved pv can also carry protocol-specific information, which we omit here.

In NOPaxos, the view $v$ is a pair $(\mathit{sessionId}, \mathit{viewId})$, where $\mathit{sessionId}$ increments when the sequencer crashes and $\mathit{viewId}$ increments when the leader replica fails. In Corfu, a Paxos-based coordinator manages views, so $v$ is a single configuration epoch. $\mathit{rids}$ is the set of replicas that participate in view $v$. It is constant in NOPaxos, whereas Corfu relies on the coordinator to reconfigure the replica set for every view, so entry pvs from different views can carry different $\mathit{rids}$.

A marker pv is resolved at one of three lifecycle events: \textsc{StartViewChange}, when the replica starts a view change; \textsc{CompleteRecovery}, when it has recovered the log entries of previous views; and \textsc{StartNormal}, when it begins normal processing in the new view. $\mathit{basePos}$ records the log position at which the new view's normal processing begins.


An entry pv is resolved whenever a replica handles a request during normal processing. The request identifier $\mathit{id}$ and the value domain of $\mathit{action}$ are protocol-specific.

\Para{NOPaxos.} A request is identified by $\mathit{id} = (\mathit{sessionId}, \mathit{seqNo})$: the session in which the sequencer multicast it and the sequence number it received. Both fields are needed because the sequencer resets $\mathit{seqNo}$ in every new session. Note that $e.\mathit{id}.\mathit{sessionId}$ may be lower than $e.v.\mathit{sessionId}$ when a request multicast in one session is committed only after the sequencer has been replaced. The $\mathit{action}$ is \texttt{Normal} or \texttt{NOOP}. A prophecy variable may contain two entry pvs for the same log position $pos$ that differ only in $\mathit{action}$, e.g., \texttt{Normal} when the request arrives and is placed at $pos$, and \texttt{NOOP} when the leader later runs gap agreement for that slot $pos$.

\Para{Corfu.} A request is identified by $\mathit{id} = (v, \mathit{seqNo})$, where $v$ is the view in which it was sequenced. Corfu's sequencer assigns globally increasing sequence numbers that are not reset across views, so a request's log position equals its sequence number. The $\mathit{action}$ is \texttt{Normal}, \texttt{Junk}, or \texttt{NOOP}, where \texttt{Junk} marks a slot that a replica filled provisionally when a request with larger $\mathit{seqNo}$ arrives first. But junk value can be later replaced by the request or finalized as a no-op, which also causes the replica's prophecy variable to hold two entry pvs for the same log position but with different  $\mathit{action}$s.

\Para{LazyLog.} Unlike in NOPaxos and Corfu, a sequencer-assigned $\mathit{seqNo}$ does not uniquely identify a request in LazyLog, because a new leader sequencer may overwrite the order established by its predecessor. \sysname{} instead uses $\mathit{id} = (\mathit{clientId}, \mathit{bootID}, \mathit{requestID})$, where $\mathit{requestID}$ is a per-client counter and $\mathit{bootID}$ is a fresh identifier that \sysname{} generates for the client on every reboot. Together with $\mathit{clientId}$, they identify a request uniquely across client reboots. The $\mathit{action}$ is \texttt{Normal} or \texttt{NOOP}. LazyLog's sequencers also replicate requests, so each sequencer owns a prophecy variable as well. A sequencer replicates only metadata and always resolves \texttt{Normal}, whereas a replica resolves \texttt{NOOP} for a position whose payload never arrives.

\subsection{Prophecy Resolution}
\label{sec:prophecy-resolution}
Prophecy resolution takes place at the replicas. A resolution records an execution outcome at a replica that the sequencer cannot know when it multicasts the request. But the sequencer-side proof can peek at these pvs to decide how to extend the ghost log. This subsection describes resolution at the replicas; \S\ref{sec:peek} describes the prophecy peek.

In all SAC systems verified by \sysname, Replica-$r$ must acquire $\mathit{permission}(r, v)$ from the coordinator through the escrow pattern (\S\ref{sec:escrow}) before it resolves any pv in view $v$. It then follows a common two-phase pattern: (1) it resolves entry pvs for the log entries recovered from previous views, and (2) it resolves entry pvs for new requests as it processes them. Marker pvs delimit these phases. The protocols differ in how requests reach the replicas and in which actions an entry pv may record.

\Para{Marker pv resolution and interaction with coordinator.} Marker pvs are resolved during a replica's interaction with the coordinator. The coordinator is built on a conventional consensus protocol such as Paxos or Raft and recovers from crashes on its own. Its internals are outside the verification scope of \sysname{}, and Grove's verified Paxos could serve as an implementation. The interaction between the coordinator and the replicas/sequencers, however, is pivotal to the correctness of SAC protocols, and \sysname{} verifies it.

The coordinator records the current view and the replicas and sequencers active in it. Replicas and sequencers interact with it through three RPCs.

\Para{\texttt{CheckRPC}.}
Replicas and sequencers periodically call \texttt{CheckRPC} on the coordinator. It doubles as a heartbeat: if a server falls silent past a timeout and its failure halts service, the coordinator increments the view identifier. The reply carries the coordinator's current view; a server that sees a higher view than its own starts a view change. Before Replica-$r$ starts the view change toward view $v$, it resolves a marker pv $m$ from $pv_r$ with $m.v = v$ and $m.\mathit{event} = \textsc{StartViewChange}$.

\Para{\texttt{PermissionRPC}.} After the view change, a replica must not resolve entry pvs for its recovered log right away, because of the risk of duplicate resolution (\S\ref{sec:at-most-once-resolution}, \S\ref{sec:escrow}). It first obtains $\mathit{permission}(r, v)$ from the coordinator through \texttt{PermissionRPC}, which implements the escrow pattern. Holding the permission, Replica-$r$ resolves an entry pv for every recovered entry in view $v$, then a marker pv $m$ with $m.v = v$ and $m.\mathit{event} = \textsc{CompleteRecovery}$, and reports its progress to the coordinator through \texttt{StablizeRPC}.

\Para{\texttt{StabilizeRPC}.} Once enough replicas and sequencers have reported ready, the coordinator tells them to resume service in view $v$. On that notification, Replica-$r$ resolves a marker pv $m$ with $m.v = v$, $m.\mathit{event} = \textsc{StartNormal}$, and $m.\mathit{basePos}$ set to the log position of the view's first new request, then resumes normal processing.

During normal processing, Replica-$r$ resolves one or more entry pvs from $pv_r$ for every request $req$ it processes. The details depend on the protocol, as described below.

\Para{NOPaxos.}
When $req$ arrives and $\mathit{req}.\mathit{id}.\mathit{seqNo}$ matches the next expected slot, Replica-$r$ appends it at log position $pos$ and resolves an entry pv $e$ from $pv_r$:
\[
e.v = v \;\land\; e.\mathit{id} = \mathit{req}.\mathit{id} \;\land\; e.\mathit{pos} = \mathit{pos} \;\land\; e.\mathit{action} = \texttt{Normal}
\]
If the leader later runs gap agreement for that position, Replica-$r$ resolves a second entry pv $e'$ from $pv_r$ that agrees with $e$ on $v$, $\mathit{id}$, and $\mathit{pos}$ but has $e'.\mathit{action} = \texttt{NOOP}$.

\Para{Corfu.}
When $req$ arrives at Replica-$r$, the replica first checks the log positions below $\mathit{req}.\mathit{id}.\mathit{seqNo}$. For every position $p < \mathit{req}.\mathit{id}.\mathit{seqNo}$ that is still empty, it fills $p$ with a junk value and resolves an entry pv $e'$ from $pv_r$:
\[
\begin{aligned}
&e'.v = v \;\land\; e'.\mathit{id} = (v, p) \;\land\; e'.\mathit{pos} = p \\
&\quad \land\; e'.\mathit{action} = \texttt{Junk} \;\land\; e'.\mathit{rids} = \mathit{rids}
\end{aligned}
\]
It then places $req$ at its own position and resolves the pv $e$:
\[
\begin{aligned}
&e.v = v \;\land\; e.\mathit{id} = \mathit{req}.\mathit{id} \;\land\; e.\mathit{pos} = \mathit{req}.\mathit{id}.\mathit{seqNo}\\
&\quad \land\; e.\mathit{action} = \texttt{Normal} \;\land\; e.\mathit{rids} = \mathit{rids}
\end{aligned}
\]
Both resolutions record $\mathit{rids}$, i.e., the replicas that form the chain in view $v$. Since the coordinator may reconfigure the chain at every view change, the sequencer-side proof needs $e.rids$ to know whose prophecy variables to examine (\S\ref{sec:peek}). Moreover, because Corfu never resets sequence numbers after a sequencer crash (NOPaxos does), a request's log position is simply its sequence number, $e.\mathit{pos} = \mathit{req}.\mathit{id}.\mathit{seqNo}$.

A filled junk value is later replaced either (1) by a \texttt{Normal} entry, if the missing request eventually arrives, or (2) by a \texttt{NOOP} entry, if the leader (head) replica synchronizes with the other replicas and freezes the position against late arrivals. Each replacement resolves a further entry pv with the new $action$, so the entry pvs capture both execution traces, from \texttt{Junk} to \texttt{Normal} and from \texttt{Junk} to \texttt{NOOP}.

\Para{LazyLog.} In LazyLog (Erwin-st), both sequencers and replicas resolve entry pvs, since both replicate request data: sequencers the metadata, replicas the payload. When $req$ arrives at a sequencer in view $v$, the sequencer appends it at its \emph{local} position $\mathit{localPos}$ and resolves an entry pv $e$:
\[
\begin{aligned}
&e.v = v \;\land\; e.\mathit{id} = \mathit{req}.\mathit{id} \;\land\; e.\mathit{pos} = \mathit{localPos}\\
&\quad \land\; e.\mathit{action} = \texttt{Normal} \;\land\; e.\mathit{rids} = \mathit{rids}
\end{aligned}
\]

Different sequencers may place $req$ at different local positions. During LazyLog's background ordering, the leader sequencer agrees with the follower sequencers and the replicas on a single position $\mathit{globalPos}$ for $req$, and each of them resolves an extra entry pv with $e.\mathit{pos} = \mathit{globalPos}$.

Sequencers hold no payload, so their entry pvs always carry $e.\mathit{action} = \texttt{Normal}$. A replica, in contrast, may discover during background ordering that it never received $req$'s payload. As in NOPaxos, the leader replica then reaches gap agreement with the followers, and the replicas resolve the entry pv with $e.\mathit{action} = \texttt{NOOP}$.

\subsection{Prophecy Peek}
\label{sec:peek}

When the sequencer processes a request, the sequencer-side proof peeks at the prophecy variables of all replicas, across views, to determine whether the request will be committed. Each protocol has its own commitment condition; the differences lie in the quorum requirement and in protocol-specific constraints. Here we explain NOPaxos' peek process in detail and describe Corfu's and LazyLog's peek in appendix(\S\ref{sec:peek-corfu}--\ref{sec:vsac-lazylog}).

\Para{NOPaxos.} When $req$ passes through the sequencer, the sequencer tags it with $req.id = (\mathit{sessionId}, \mathit{seqNo})$. The sequencer-side proof decides whether $req$ will eventually be committed, and further decides whether it can extend the ghost log. We formalize this behavior as the \emph{logically atomic Hoare triple}~\cite{popl19-prophecy-variable} in Figure~\ref{fig:sequencer-spec}.
\begin{figure}[!t]
\begin{tcolorbox}[colback=white!10,
                  width=\columnwidth,
                  arc=1mm, auto outer arc,
                  boxrule=0.5pt,
                  left=1pt, right=1pt, top=2pt, bottom=2pt, boxsep=0pt,
                 ]
\begin{lstlisting}[style=rocqspec,aboveskip=0pt,belowskip=0pt]
⊢ {{{ is_sequencer s state γ N f ∗ ...}}}
  <<< ∀∀ (ghost_log : loghist), ghost_log_half γ ghost_log >>>
    Sequencer__handleMessage #s  @ ↑nopaxosN
  <<< ∃∃ (committed : bool) (et : log_entry_type),
      ghost_log_half γ (if committed then ghost_log ++ [et] 
      			else ghost_log) >>>
  {{{ RET #(); ... }}}.
\end{lstlisting}
\end{tcolorbox}
\caption{The logically atomic Hoare triple for the NOPaxos sequencer's message handler.}
\Description{A logically atomic Hoare triple specifying the NOPaxos sequencer's message handler.}
\label{fig:sequencer-spec}
\end{figure}
The angle brackets \lstinline[style=rocqspec,basicstyle=\small\ttfamily]|<<< >>>| delimit the atomic update at linearization point: the
handler either appends an entry to ghost log or leaves it unchanged. To make the decision, the proof distinguishes two cases.

\Para{Case 1: $req$ is committed during normal processing.} After the sequencer multicasts $req$, the request is committed without any intervening sequencer or replica failure, that is, without a view change. The commit condition is
\[
\begin{aligned}
	&\texttt{Commit}_1(req) \;\triangleq\; \\
	&\quad\exists\ m: m.v.sessionId = req.id.sessionId  \\
	&\qquad\land\ \exists\ Q_m: \texttt{qsMarker}(Q_m, m, \textsc{StartNormal}) \\ 
	&\qquad\land\ \exists\ e: e.v=m.v \land e.id = req.id \land\texttt{qsPrefix}\ (e)
\end{aligned}
\label{eq:normal-quorum}
\]

We explain $\texttt{Commit}_1(req)$ in more detail. The sequencer-side proof performs a two-step check during the peek.

(1) A leader-inclusive quorum of $f+1$ replicas has started normal processing in some view, so their prophecy variables all contain the same marker pv $m$ with $m.\mathit{event} = \textsc{StartNormal}$. We define this check as $\texttt{qsMarker}(Q, m, \mathit{ev})$, and a leader-inclusive quorum $Q$ that supports a pv $p$ as $\texttt{qSupport}(Q, p)$, so $\texttt{qSupport}(Q, m)$ implies that a quorum of replicas have started normal processing in the view $m.v$.
\[
\begin{aligned}
&\texttt{qsMarker}(Q, m, \mathit{ev}) \triangleq m.\mathit{event} = \mathit{ev} \land \texttt{qSupport}(Q, m)
\end{aligned}
\]
\[
\begin{aligned}
&\texttt{qSupport}\ (Q, p) \;\triangleq\; Q \subseteq \{0, \ldots, 2f\} \;\land\; |Q| \geq f+1 \\
&\qquad \land\; p.v.\mathit{viewId} \bmod (2f+1) \in Q \;\land\; \forall r \in Q:p \in pv_r
\end{aligned}
\]

(2) Given the candidate view $e.v$, the proof checks whether the pvs provide sufficient evidence that $req$ is committed: a leader-inclusive quorum of replicas must hold an entry pv $e$ for $req$ in this view.  Besides, $req$ can commit only after every request multicast before it in the same session has been committed in sequence-number order. The proof therefore also checks that the replicas' prophecy variables support both $req$ and $req$'s prefix, defined as $\texttt{qsPrefix}(e)$:
\[
\begin{aligned}
&\texttt{qsPrefix}\ (e) \;\triangleq\; \forall\ 1\leq \mathit{sn} \leq \mathit{e}.\mathit{id}.\mathit{seqNo}:\ \\ 
&\quad\exists\ e': e'.v=e.v \land e'.id.\mathit{seqNo}=sn  \\
&\quad\quad\  \land e.pos- e'.pos = e.id.\mathit{seqNo} - e'.id.\mathit{seqNo}  \\
&\quad\quad\ \land  \exists\ Q':\ \texttt{qSupport}(Q', e') 
\end{aligned}
\]

\Para{Case 2: $req$ is committed during recovery.} Alternatively, $req$ is not committed right after the multicast: the sequencer or a replica crashes and triggers a view change, so $\mathit{sessionId}$ or $\mathit{viewId}$ is incremented. After the view change, $req$ is included in the recovered log and replicated to a leader-inclusive quorum, which also commits it. The commit condition is
\[
\begin{aligned}
	&\texttt{Commit}_2(req) \;\triangleq\; \\
	&\quad\exists\ m: m.v.\mathit{sessionId} \geq req.id.\mathit{sessionId} \\
	&\qquad\land\ \exists\ Q_m: \texttt{qsMarker}(Q_m, m, \textsc{CompleteRecovery}) \\
	&\qquad\land\ \exists\ e: e.v = m.v \land e.id = req.id \\
	&\qquad\qquad\land\ \texttt{qsRecoverPrefix}\ (Q_m, e)
\end{aligned}
\]

\[
\begin{aligned}
&\texttt{qsRecoverPrefix}\ (Q, e) \;\triangleq\; \forall\ 1\leq \mathit{sn} \leq \mathit{e}.\mathit{id}.\mathit{seqNo}:\ \\ 
&\quad\exists\ e': e'.v=e.v \land e'.id.\mathit{seqNo}=sn  \\
&\quad\quad\  \land e.pos- e'.pos = e.id.\mathit{seqNo} - e'.id.\mathit{seqNo}  \\
&\quad\quad\ \land  \texttt{qSupport}(Q, e') 
\end{aligned}
\]

The proof considers $req$ committed if  replicas' prophecy variables provide sufficient evidence for either case:
\[
\texttt{Commit}\ (req) \;\triangleq\; \texttt{Commit}_1\ (req) \vee \texttt{Commit}_2\ (req)
\]

A committed $req$ lets the sequencer-side proof atomically extend the ghost log with an entry, depending on $e.\mathit{action}$ in the commit condition that $req$ satisfies. If the quorum of replicas agree on $e.\mathit{action} = \texttt{Normal}$, the ghost log is appended with a \texttt{Normal} entry, meaning that $req$ is committed with its payload. Otherwise, it is appended with a \texttt{NOOP} entry, meaning that $req$'s log position is finally marked as a no-op.

If no such evidence exists, $req$ is never committed in this execution, and the ghost log remains unchanged.

\Para{Corfu.} The peek for Corfu follows the same shape as for NOPaxos: the sequencer-side proof checks whether $req$ commits during normal processing or during recovery. There are two major differences. (1) Corfu commits a request only when every replica in the chain acknowledges it, so the proof must find an entry pv for $req$ in \emph{every} replica's prophecy variable, not just a majority. (2) Corfu is a shared log, and an append is acknowledged without executing the entries before it, so the normal-processing condition $\texttt{Commit}_1(req)$ does not require the replicas' prophecy variables to agree on the prefix of earlier requests. Appendix~\ref{sec:peek-corfu} explains more details.

\Para{LazyLog.} The peek for LazyLog follows the same shape as for Corfu, with two major differences. (1) LazyLog's sequencers replicate request metadata, so the full quorum that the proof examines includes every sequencer as well as every replica. (2) The leader sequencer decides only a tentative order: $req$ may commit at the log position $pos$ that the leader assigns, or at a different position that a new leader decides after a view change. The proof therefore also checks the leader's \emph{scope}. It peeks at the subsequent marker pvs to learn the last log position committed by the current leader. If $pos$ goes beyond this scope, extending the ghost log for $req$ is the responsibility of a later leader, not the current one. Otherwise, the proof follows conditions similar to Corfu's to decide whether $req$ is committed and, if so, whether it is committed as \texttt{Normal} or \texttt{NOOP}. Appendix~\ref{sec:vsac-lazylog} explains more details.

\section{Implementing Verified SAC Systems}
\label{sec:implement-verified-protocol}

%

\Para{Codebase.} To implement the verified versions of SAC systems, we take the open-source implementations of NOPaxos~\cite{nopaxos-implementation}, Corfu~\cite{corfu-implementation} and LazyLog~\cite{lazylog-implementation} as reference and port them to Go, and translate the Go code into Rocq specifications with Goose~\cite{goose}. Based on the specification, we develop machine-checked proofs on top of them in Grove~\cite{sosp23-grove,grove-thesis}. The open-source code is incomplete with respect to failure handling: NOPaxos implements view change but not replica rejoin; Corfu and LazyLog implement neither view change nor most of the crash-recovery logic. For the missing parts we follow the protocol descriptions in the respective papers. Verification then iterates: when a proof attempt exposes a bug, we revise the implementation, translate it again, and resume the proof, until the implementation is fully verified. Coding agents such as Codex and Claude provide some assistance to develop the proof, but the key proof insights still require human expertise to audit, revise, and supply. Table~\ref{tab:code-size} summarizes the size of each codebase. 




\begin{table}[!t]
	\centering
	\footnotesize
	\setlength{\tabcolsep}{2.5pt}
	\caption{Code size (lines of code) of SAC protocols}
	\label{tab:code-size}
	\begin{tabular}{lccccc}
		\toprule
		\textbf{Protocol} &
		\makecell{\textbf{Open-Source}\\\textbf{Code}} &
		\makecell{\textbf{Verified}\\\textbf{Code}} &
		\makecell{\textbf{Rocq}\\\textbf{Spec}} &
		\makecell{\textbf{Rocq}\\\textbf{Proof}} &
		\makecell{\textbf{Proof/Source}\\\textbf{Ratio}} \\
		\midrule
		NOPaxos & 4,914 & 8,442 & 7,810 & 175,808 & 20.8$\times$ \\
		Corfu   & 2,365 & 5,830 & 5,847 & 86,305  & 14.8$\times$ \\
		LazyLog & 3,552 & 7,363 & 7,856 & 183,239 & 24.9$\times$ \\
		\bottomrule
	\end{tabular}
\end{table}

\Para{Trusted code and axioms.} Our verification relies on several trusted assumptions, which appear as axioms in the proofs. (1) We assume that network communication is implemented correctly, so we reuse the network axioms built into Grove and focus on the protocol logic. (2) We assume that universally unique identifiers can be generated~\cite{disc17-dcr-techreport}. Our implementation obtains them through a trusted API, \texttt{getUUID}, and one corresponding axiom supports both the proof of the escrow pattern (\S\ref{sec:escrow}) and the fix for the client-reboot bug we found in both protocols (\S\ref{sec:other-bugs}). (3) The coordinator's crash recovery is outside the scope of \sysname{}, and Grove already provides a verified Paxos module that could serve this role. 


\Para{Public APIs and verified spec.} Each SAC protocol exposes the two public APIs in Figure~\ref{fig:sac-api}, and each API carries the logically atomic Hoare triple in Figure~\ref{fig:client-spec} that specifies its behavior and is verified in Grove. When a write succeeds, the ghost log holds the corresponding entry at some position $\mathit{gpos}$, and the reply carries that position. The postcondition also hands the client a witness that the entry is recorded there, which the client can later pass to the read API. A successful read returns that entry, with a witness that it is the one the ghost log holds at that position.

\begin{figure}[!t]
\begin{tcolorbox}[colback=white!10,
	width=\columnwidth,
	arc=1mm, auto outer arc,
	boxrule=0.5pt,
	left=3pt, right=1pt, top=2pt, bottom=2pt, boxsep=0pt,
	]
\begin{lstlisting}[style=apibox]
// Submit req; returns a reply once req commits or errors.
// On commit, the reply carries req's committed log position.
WriteReply write(Request req);
// Read position pos; returns the entry committed at pos.
ReadReply read(int pos);
\end{lstlisting}
\end{tcolorbox}
\caption{\sysname public APIs.}
\Description{The two public API signatures that every verified SAC protocol exposes to applications.}
\label{fig:sac-api}
\vspace{-0.3cm}
\end{figure}

\begin{figure}[!t]
\begin{tcolorbox}[colback=white!10,
	width=\columnwidth,
	arc=1mm, auto outer arc,
	boxrule=0.5pt,
	left=1pt, right=1pt, top=2pt, bottom=1pt, boxsep=0pt,
	]
	\begin{lstlisting}[style=rocqspec,aboveskip=0pt,belowskip=0pt]
⊢ {{{ is_client c state γ N f ∗ own_client c state γ requestID ∗
    is_request req r }}}
  <<< ∀∀ (ghostLog : loghist), ghost_log_half γ ghostLog >>>
    Client__Write #c #req @ ↑vsacN
  <<< ∃∃ (success : bool) (gpos : u64),
        ghost_log_half γ ghostLog ∗
        ⌜success = true → ghostLog !! gpos = Some (entry_of r)⌝ >>>
  {{{ (replyLoc : loc), RET (#replyLoc, #success);
        own_client c state γ (requestID + 1) ∗
        (⌜success = false⌝ ∨
         ⌜success = true⌝ ∗
          replyLoc ↦[WriteReply::"LogPosition"] #gpos ∗
          server_log_witness γ gpos (entry_of r)) }}}.

⊢ {{{ is_client c state γ N f ∗ own_client c state γ requestID }}}
  <<< ∀∀ (ghostLog : loghist), ghost_log_half γ ghostLog >>>
    Client__Read #c #posW @ ↑vsacN
  <<< ∃∃ (success : bool) (et : log_entry_type),
        ghost_log_half γ ghostLog ∗
        ⌜success = true → ghostLog !! posW = Some et⌝ >>>
  {{{ (replyLoc : loc), RET (#replyLoc, #success);
        own_client c state γ (requestID + 1) ∗
        (⌜success = false⌝ ∨
         ⌜success = true⌝ ∗
          replyLoc ↦[ReadReply :: "LogPosition"] #posW ∗
          replyLoc ↦[ReadReply :: "Entry"] (log_entry_to_val et) ∗
          server_log_witness γ posW et) }}}.
	\end{lstlisting}
\end{tcolorbox}
\caption{Logically atomic Hoare triples for \sysname's client APIs. We omit several conjuncts for readability.}
\Description{Logically atomic Hoare triples for the client write and read APIs.}
\label{fig:client-spec}
\end{figure}

\section{Other Bugs and Lessons}
\label{sec:other-bugs}

Besides the duplicate-resolution bug in \S\ref{sec:at-most-once-resolution}, our verification with \sysname uncovered several other bugs. The first two are safety bugs. In both cases the  proof could not be completed, and the failed proof obligation led us to the bug. The third is a liveness bug in our own implementation, which the proof cannot detect and which we found during testbed evaluation.

\Para{Client reboot.}
The open-source implementations of all the three protocols mishandle client reboots. A client maintains an in-memory counter $\mathit{requestID}$ to uniquely identify the requests it submits. The counter is volatile and its value is lost when the client crashes. After rebooting, the client may reuse a $\mathit{requestID}$ for a different request. If a reply to the earlier request is delayed in the network and arrives after the reboot, the client mistakes it for a reply to the new request and wrongly considers that request committed. Our fix is inspired by crash vectors~\cite{disc17-dcr-techreport}: on every reboot, the client generates a fresh universally unique identifier (UUID) $\mathit{bootID}$ and attaches it to each request alongside $\mathit{requestID}$. The pair names a request uniquely across reboots, and replicas echo both in their replies (Figure~\ref{fig:client-spec}), so the client can tell stale replies apart from replies to its current request.

\Para{Premature truncation of logs.}
The second safety bug lies in NOPaxos's view change. In NOPaxos's open-source implementation~\cite{nopaxos-implementation}, the replica attempts to truncate its log prematurely during view change. When its last normal view is below the reconstruction pivot in \textsc{Start-View}, the replica rewinds to its local commit point before fetching the replacement suffix. This rewind can target already executed entries even when the existing log agrees with the recovered log. Because rollback is unimplemented, the replica aborts instead of completing recovery. We identified an execution with five replicas and two external crashes in which this premature rewind aborts a third replica, leaving fewer than a quorum alive and preventing progress even after communication recovers. Our fix defers truncation until the suffix arrives and the combined log is validated against the \textsc{Start-View} summary, so an interrupted transfer always leaves an intact log. Since the bug lives entirely in this truncation optimization, model checking the high-level protocol spec cannot find it.

\Para{Liveness bug.}
We also encountered a liveness bug in our verified implementation of NOPaxos. \sysname{} verifies safety only, so the proof checker cannot detect such bugs, and we found this one during testbed evaluation. The bug arises when the leader replica instructs followers to start a new view. A follower whose log is inconsistent with the leader's log issues a state transfer request to fetch the missing entries. Our implementation imposed an overly strict validity check on these requests and rejected all of them. The followers therefore never obtain the missing entries and remain in view-change state, while the leader never collects enough responses to enter the new view. The system stops making progress. Because the bug does not affect safety, its proof still passed the proof checker. We fixed the bug and re-ran the proof checker against the corrected code. This bug illustrates a limitation of \sysname{}: it proves linearizability but not liveness, so testing remains necessary to catch progress bugs.

\section{Evaluation}

We evaluate the performance of our verified implementations against their open-source, unverified counterparts. All experiments run on Google Cloud, using \texttt{n1-standard-4} VMs for clients and \texttt{n1-standard-16} VMs for sequencers, replicas, and coordinators. The open-source implementation of Corfu~\cite{corfu-implementation} and LazyLog~\cite{lazylog-implementation} communicates over RDMA, which is not supported by Google Cloud. For a fair comparison, we replace their communication library, eRPC~\cite{nsdi19-erpc}, with rpclib~\cite{rpclib}, so that the verified and unverified versions both communicate over TCP. We refer to our verified implementations as \NOPaxosV, \CorfuV, \LazyLogV, respectively. For each test case, we run 5 trials and report the median value among them.

\begin{figure}[!t]
    \centering
    \includegraphics[width=0.269760843\textwidth]{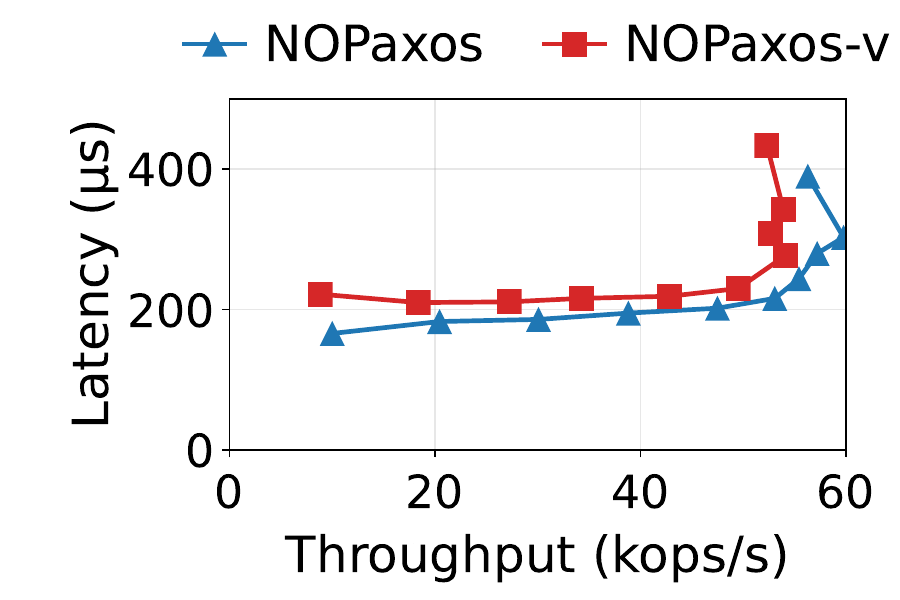}
    \caption{NOPaxos vs.\ \NOPaxosV}
    \Description{Latency versus throughput for NOPaxos and \NOPaxosV during normal processing.}
    \label{fig:nopaxos-latency-throuput}
\end{figure}

\begin{figure}[!t]
    \centering
    \includegraphics[width=0.3\textwidth]{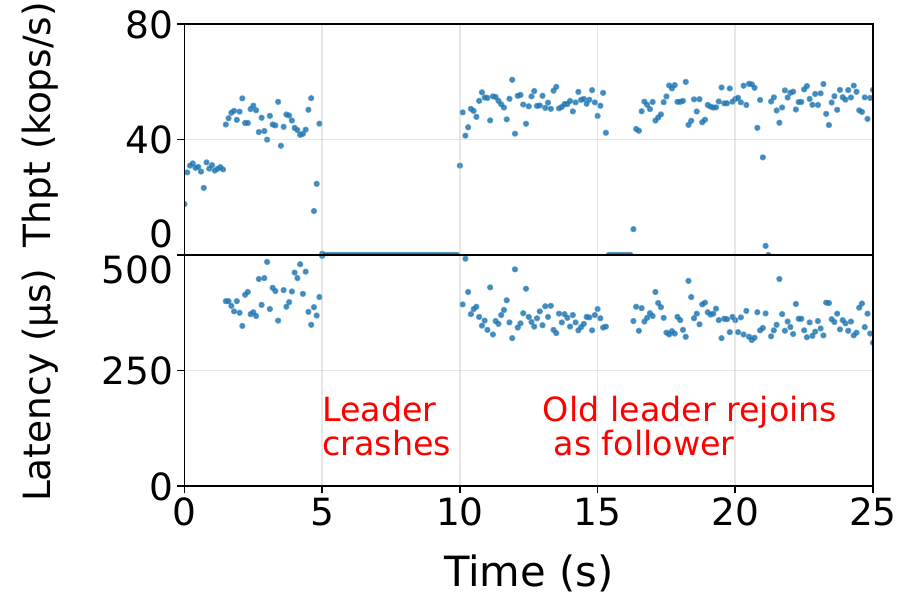}
    \caption{\NOPaxosV crash recovery}
    \Description{Stacked throughput and latency timelines sharing a time axis during \NOPaxosV leader crash and rejoin.}
    \label{fig:nopaxos-crash-recovery}
\end{figure}

\Para{\NOPaxosV.} We deploy NOPaxos and \NOPaxosV with 3 replicas each. Figure~\ref{fig:nopaxos-latency-throuput} shows their throughput and median latency during normal processing as we keep adding closed-loop clients. \NOPaxosV performs comparably to NOPaxos, with a gap within 15\%. Figure~\ref{fig:nopaxos-crash-recovery} shows \NOPaxosV under a crash and recovery. We fix the load at 20 clients, which brings \NOPaxosV close to saturation, and kill the leader at $t = 5$\,s. Successful client requests resume about 5 seconds after the crash. At $t = 15$\,s, we restart the failed replica. It rejoins as a follower, which triggers another view change, and client requests resume about 1.4 seconds after restart. This recovery time is comparable to the numbers reported from state-of-the-art consensus protocols in public cloud~\cite{vldb23-nezha,sosp25-tiga}.

\begin{figure*}[!t]
    \centering
    \begin{minipage}[t]{0.66\textwidth}
        \vspace{0pt}
        \centering
        \begin{subfigure}[t]{0.48485\linewidth}
            \centering
            \includegraphics[width=0.843\linewidth]{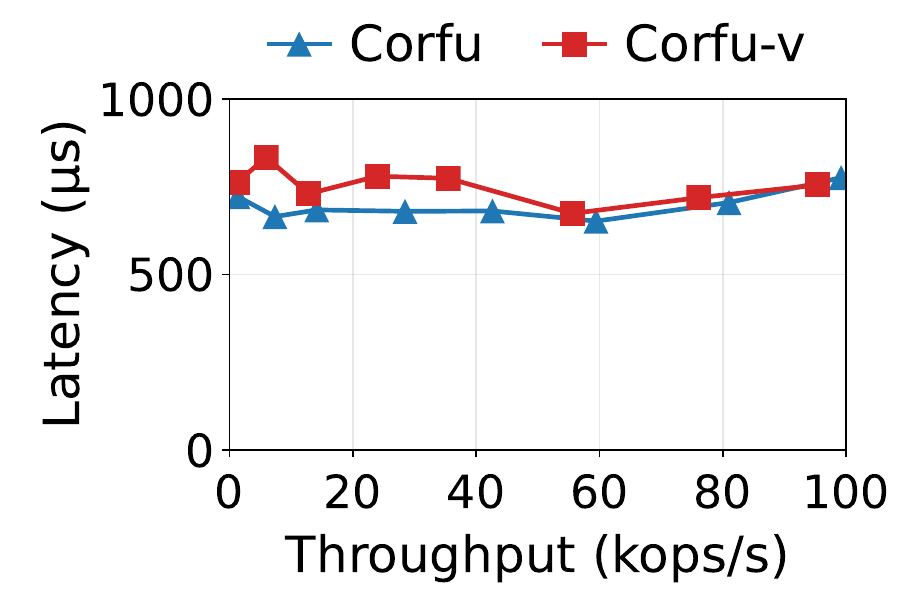}
            \caption{Append-only}
            \label{fig:corfu-latency-thpt}
        \end{subfigure}\hfill
        \begin{subfigure}[t]{0.48485\linewidth}
            \centering
            \includegraphics[width=0.843\linewidth]{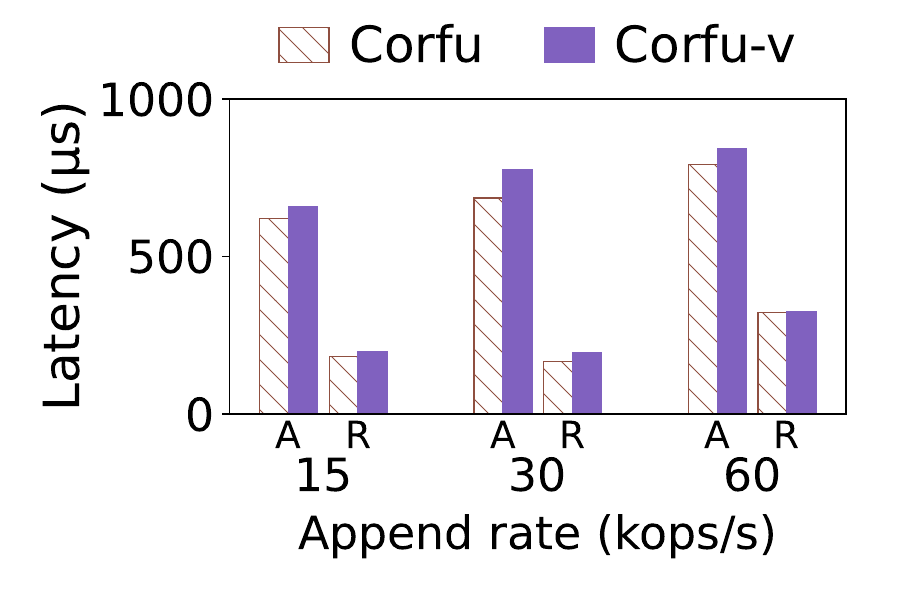}
            \caption{Append+read (A: append, R: read)}
            \label{fig:corfu-mixed-append-read-latency}
        \end{subfigure}
        \caption{Corfu vs.\ \CorfuV}
        \Description{Latency versus throughput for an append-only workload, and latency for a mixed append and read workload.}
        \label{fig:corfu-steady}
    \end{minipage}\hfill
    \begin{minipage}[t]{0.32\textwidth}
        \vspace{0pt}
        \centering
        \includegraphics[width=\linewidth]{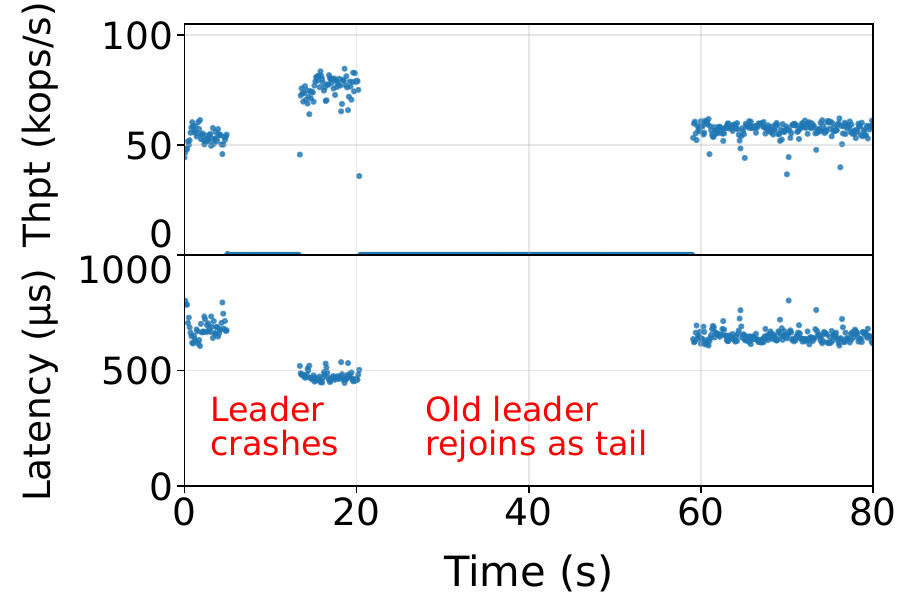}
        \caption{\CorfuV crash recovery}
        \Description{Stacked throughput and latency timelines sharing a time axis during \CorfuV crash recovery.}
        \label{fig:corfu-crash-recovery}
    \end{minipage}
\end{figure*}

\Para{\CorfuV.} We deploy Corfu and \CorfuV with a chain of 3 replicas and compare them under two workloads: appends only (Figure~\ref{fig:corfu-latency-thpt}), and appends and reads issued concurrently (Figure~\ref{fig:corfu-mixed-append-read-latency}). Every append carries a 4KB payload. Under the append-only workload, we increase the client number from 1 to 60. \CorfuV adds 2.0\%--14.7\% to append latency and gives up 3.9\%--16.9\% of throughput.

In the mixed workload, each client runs a writer thread and a reader thread. The writer appends at a fixed rate of 1K ops/s and the reader repeatedly reads the latest entry its writer has committed, so 15, 30, and 60 clients drive append rates of 15, 30, and 60 kops/s. A read is served by the tail replica in a single round trip rather than traversing the chain, which is why read latency sits well below append latency. Compared with Corfu, \CorfuV adds 6.1\%--13.1\% to append latency and 0.9\%--16.2\% to read latency.

We evaluate \CorfuV during crash recovery using 40 append-only clients (Figure~\ref{fig:corfu-crash-recovery}). When the head of the chain fails, the coordinator shrinks the chain from 3 replicas to 2. Appends then traverse a shorter chain, so latency drops. Recovery is slower than in NOPaxos for two reasons. Appends carry 4KB payloads rather than the 100--200 bytes NOPaxos replicates, and \CorfuV has no snapshot support, so a rejoining replica must fetch every log entry from scratch. This catch-up takes up to 40 seconds. Once it completes, the chain regains its original length and latency returns to its initial level.

\begin{figure*}[!t]
    \centering
    \begin{minipage}[t]{0.66\textwidth}
        \vspace{0pt}
        \centering
        \begin{subfigure}[t]{0.48485\linewidth}
            \centering
            \includegraphics[width=0.843\linewidth]{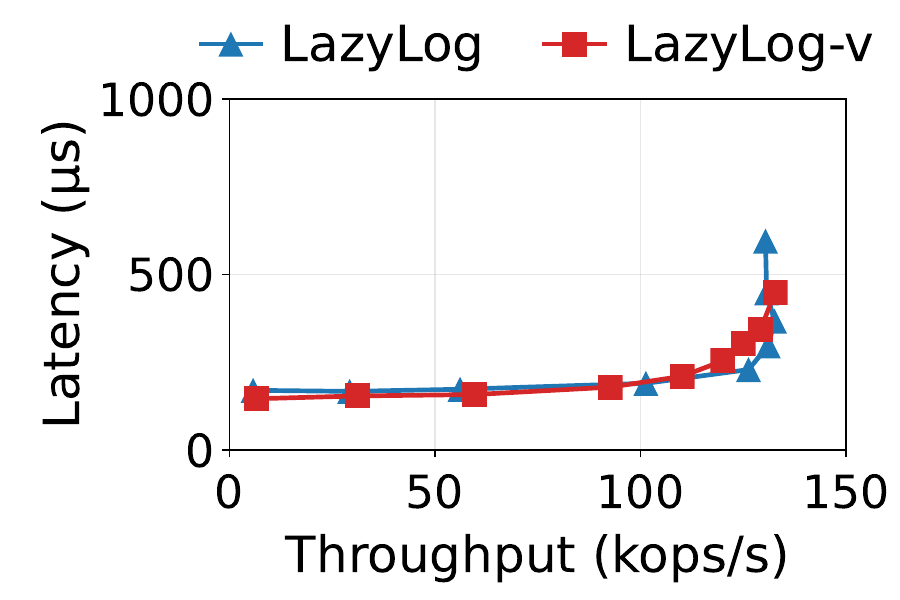}
            \caption{Append-only}
            \label{fig:lazylog-latency-thpt}
        \end{subfigure}\hfill
        \begin{subfigure}[t]{0.48485\linewidth}
            \centering
            \includegraphics[width=0.843\linewidth]{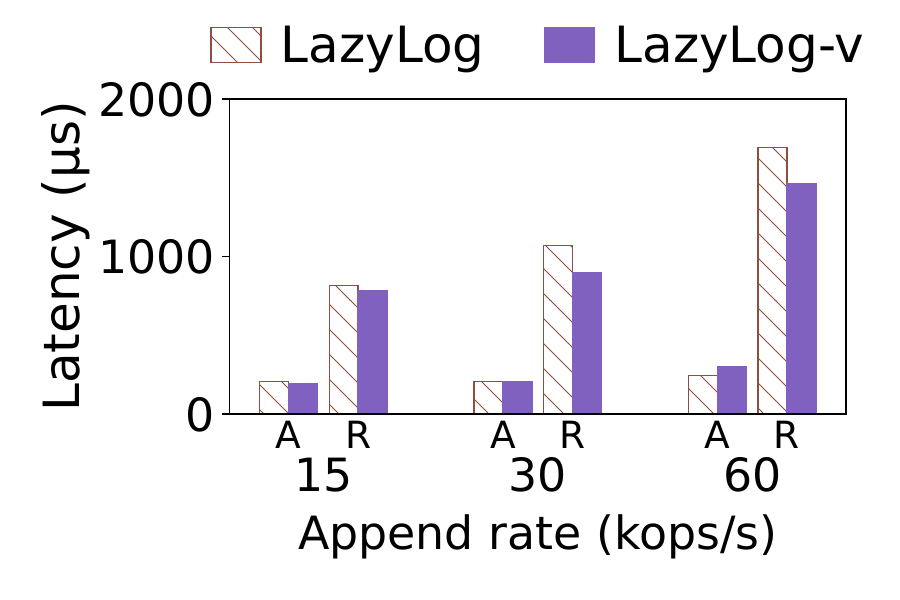}
            \caption{Append+read (A: append, R: read)}
            \label{fig:lazylog-mixed-append-read-latency}
        \end{subfigure}
        \caption{LazyLog vs.\ \LazyLogV}
        \Description{Latency versus throughput for an append-only workload, and latency for a mixed append and read workload.}
        \label{fig:lazylog-steady}
    \end{minipage}\hfill
    \begin{minipage}[t]{0.32\textwidth}
        \vspace{0pt}
        \centering
        \includegraphics[width=\linewidth]{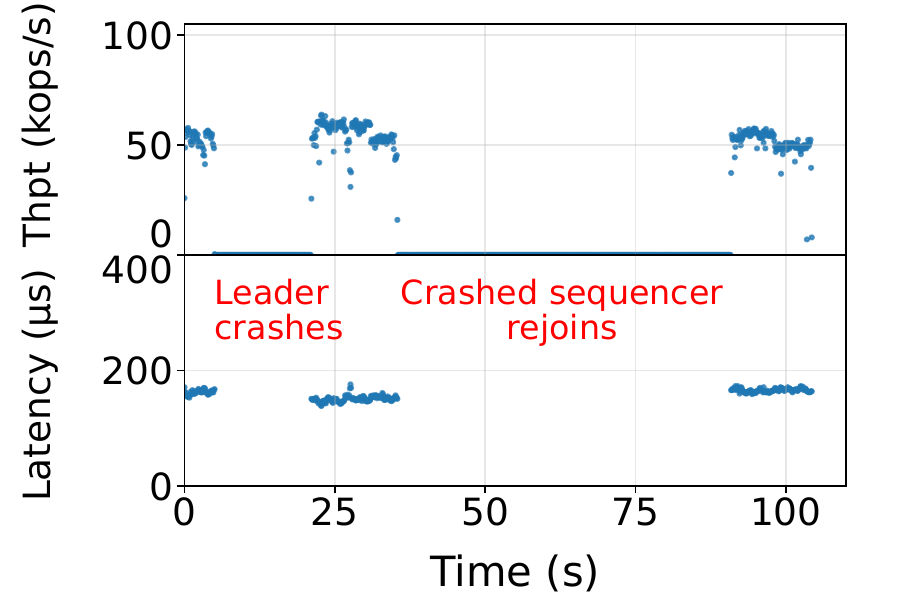}
        \caption{\LazyLogV crash recovery}
        \Description{Stacked throughput and latency timelines sharing a time axis during \LazyLogV crash recovery.}
        \label{fig:lazylog-crash-recovery}
    \end{minipage}
\end{figure*}

\Para{\LazyLogV.} We deploy LazyLog and \LazyLogV with 3 sequencers and 1 shard, and compare them under the same two workloads as Corfu. LazyLog (Figure~\ref{fig:lazylog-steady}) achieves much lower append latency than Corfu (Figure~\ref{fig:corfu-steady}), because it lazily binds records to their global linearizable order. The trade-off is higher read latency, because a read must wait until its target log position becomes stable. Between the two LazyLog implementations, \LazyLogV's append latency stays very close to LazyLog's as we increase the number of clients (Figure~\ref{fig:lazylog-latency-thpt}), and its saturation throughput at 50 clients is within 5.8\% of LazyLog's.

Under the mixed workload (Figure~\ref{fig:lazylog-mixed-append-read-latency}), \LazyLogV's read latency is even lower than LazyLog's. This gap mainly stems from how the two implementations perform background ordering. The open-source implementation delegates it to a separate server (\texttt{cons\_svr} in~\cite{lazylog-implementation}), which fetches metadata from the leader sequencer and then synchronizes with the replicas. This indirection lengthens the background-ordering path and delays the advance of the stable point (\emph{stable-gp}), which in turn adds latency to reads. \LazyLogV instead follows the design in the LazyLog paper (Figure~\ref{fig:lazylog-workflow}), in which the leader sequencer performs background ordering itself, and thus achieves lower read latency.

We evaluate \LazyLogV during crash recovery using 10 append-only clients (Figure~\ref{fig:lazylog-crash-recovery}), which drive about the same throughput as the 40 clients in the \CorfuV experiment (Figure~\ref{fig:corfu-crash-recovery}). In \CorfuV, a shorter chain lowers latency. In \LazyLogV, an append always commits in a single round trip, so latency stays nearly flat across the crash and the rejoin. Recovery, however, is slower than in \CorfuV. Because LazyLog keeps request metadata on the sequencers and payload data on the replicas, its recovery requires more complex synchronization between them. This synchronization adds several seconds to both the recovery time and the rejoin time.

\section{Discussion}


\Para{Language support.}
\sysname{} currently supports Go programs via Goose~\cite{goose} and Grove~\cite{sosp23-grove}, which covers the languages used by our target systems. However, Goose does not yet support the full Go language. For example, it does not support returning directly from within a \texttt{for} loop, and it does not handle short-circuit evaluation in \texttt{if} conditions gracefully, which can lead to branch explosion. Improving Goose's Go coverage would make it easier to develop verified systems that achieve both high performance and strong assurance. More broadly, we believe it is possible to develop translators from other languages (e.g., P4, Rust) into Grove, which would allow \sysname{} to apply to a wider range of systems.


\Para{Beyond linearizability.}
\sysname{} targets strong consistency (linearizability), but many practical systems provide weaker consistency guarantees that are equally subtle and deserve rigorous verification. For example, distributed snapshot protocols~\cite{tocs85-chandy-lamport-snapshot,vldb17-flink-snapshot} are typically concerned with serializability, and CockroachDB describes its consistency model as lying between serializability and strict serializability~\cite{CockroachDB-consistency}. We believe \sysname{} can be extended to verify such weaker consistency properties, and we leave this direction to future work.

\section{Related Work}

\Para{SAC systems.}
The complexity and performance demands of coordination systems have motivated decoupling ordering from durability, leading to a broad class of sequencer-aided coordination (SAC) systems. We demonstrate \sysname{} on Corfu~\cite{nsdi12-corfu,tocs13-corfu}, NOPaxos~\cite{osdi16-nopaxos} and LazyLog~\cite{sosp24-lazylog}, but many others in this family could benefit. NeoBFT~\cite{sigcomm23-neobft} and SwitchBFT~\cite{nsdi26-switchbft} accelerate BFT consensus with Byzantine-resilient sequencers, and their linearizability proofs fit \sysname{} naturally. Eris~\cite{sosp17-eris} and Hydra~\cite{nsdi23-hydra} extend the design of NOPaxos and should also be within reach of \sysname, with additional refinement-based reasoning~\cite{dijkstra1972-refinement-notes,back2012refinement,sosp15-ironfleet}. Timestamp-based protocols~\cite{vldb23-nezha,conext20-domino,sosp25-tiga}, which order requests by non-consecutive timestamps, can also be viewed as SAC systems; verifying them is an important direction for future work.

\Para{Formal verification frameworks.}
\sysname{} builds on a line of Rocq (Coq)~\cite{rocq9} verification frameworks. Iris~\cite{iris} provides higher-order concurrent separation logic; Perennial~\cite{sosp19-perennial} extends Iris with crash reasoning and incorporates Goose~\cite{goose} to verify Go programs; and Grove~\cite{grove-thesis,sosp23-grove} adds distributed-systems primitives such as RPC, replicated state machines, and leases. We base \sysname{} on Grove because it offers native support for these components and prophecy variables. \sysname{}'s methodology is general and could in principle be applied to other verification frameworks, including Dafny~\cite{dafny}, IronFleet~\cite{sosp15-ironfleet}, Verdi~\cite{pldi15-verdi}, and Verus~\cite{oopsla23-verus}.


\Para{Verified systems.}
Machine-checked proofs have been applied to realistic storage systems, including file systems (FSCQ~\cite{sosp15-fscq}, DFSCQ~\cite{sosp17-dfcq}), journaling (GoJournal~\cite{osdi21-gojournal}), and key-value stores (PoWER~\cite{osdi25-power}). These works reason about crash safety in a disk-based, single-node setting. \sysname{} differs in that it targets replication-based crash recovery across a broad class of distributed SAC systems.

\Para{Prophecy variables in verification.}
Prophecy variables were introduced by Abadi and Lamport~\cite{lics88-abadi-lamport,tcs91-abadi-lamport} and formalized in separation logic by Jung et al.~\cite{popl19-prophecy-variable}. They have since been used to verify local and shared-memory concurrent programs~\cite{popl20-spygame,cpp21-msqueue,vmcai24-persistent-lib}, but not crash-recoverable distributed protocols. The two prior works most closely related to \sysname{} are vMVCC~\cite{osdi23-vmvcc}, which uses prophecy variables to verify multi-version concurrency control but assumes crash-free servers, and PoWER~\cite{osdi25-power}, which uses prophecy variables for crash consistency but under a disk-based failure model rather than replication. In contrast, \sysname{} uses prophecy variables to reason about crash recovery in replicated coordination protocols via cross-view prophecy peek, extending the technique to a broader class of coordination systems.

\section{Conclusion}
\vspace{-0.2cm}
Modern coordination systems demand both high performance and strong correctness guarantees. Sequencer-aided coordination (SAC) protocols achieve higher performance than conventional protocols such as Paxos and VR, but they have lacked rigorous proofs of correctness. \sysname{} fills this gap. It introduces cross-view prophecy peek to resolve the non-determinism that crash recovery brings to the verification of SAC systems, and it adopts the escrow technique to rule out duplicate prophecy resolution during recovery. We applied \sysname{} to NOPaxos, Corfu and LazyLog, and uncovered several bugs in their open-source implementations. We further built verified implementations for all three protocols, which come with machine-checked linearizability proofs and yield comparable performance to their unverified counterparts.

\bibliographystyle{plain}
\bibliography{ref}

\clearpage
\appendix
\section{Prophecy Peek in Corfu}
\label{sec:peek-corfu}

Corfu's replica set changes across views: at each view change, the coordinator chooses the replicas that serve in the next view. To decide whether $req$ commits in view $v$, the sequencer-side proof must therefore examine the prophecy variables of exactly the replicas of $v$, which is why every entry pv carries $\mathit{rids}$ (\S\ref{sec:pv-design}).

\Para{Case 1: $req$ is committed during normal processing.} The proof considers $req$ committed in view $v$ once every replica in the chain of $v$ has processed it. The replicas need not agree on the action: some may have processed $req$ as a \texttt{Normal} entry, while others, which never received it, filled its slot with \texttt{Junk}. As long as every replica holds an entry pv for $req$'s slot, the proof can extend the ghost log with either a \texttt{Normal} entry or a \texttt{NOOP} entry.
\[
\begin{aligned}
	&\texttt{Commit}_1(req) \;\triangleq\; \\
	&\quad\exists\ e: e.v = req.id.v \\
	&\qquad\land\ \exists\ m: \texttt{allMarker}(e.\mathit{rids}, m, \textsc{StartNormal}) \\
	&\qquad\qquad\ \ \land\ \texttt{allProcessed}\ (e.\mathit{rids}, m, req.id.\mathit{seqNo})
\end{aligned}
\]
\[
\begin{aligned}
&\texttt{allMarker}(Q, m, \mathit{ev}) \triangleq \\
&\quad (m.\mathit{event} = \mathit{ev}) \land (\forall r\in Q: m \in pv_r)
\end{aligned}
\]
\[
\begin{aligned}
&\texttt{allProcessed}(Q, m, pos) \triangleq \quad \forall\ 1 \leq \mathit{sn} \leq pos: \\
&\qquad \forall r \in Q:\ \exists\ e \in pv_r:\ e.v = m.v \land e.id = (m.v, \mathit{sn})
\end{aligned}
\]

The recorded actions determine the entry. (1) If the proof can find a view $v$ where the pvs of all replicas record $req$ as \texttt{Normal}, the proof appends a \texttt{Normal} entry. (2) Otherwise, every replica's pv records $req$ but no view has all replicas agreeing on \texttt{Normal}, and the proof appends a \texttt{NOOP} entry.

\Para{Case 2: $req$ is committed during recovery.} As in NOPaxos, $req$ and every request before it must be fully replicated across the replicas of the new view:
\[
\begin{aligned}
	&\texttt{Commit}_2(req) \;\triangleq\; \\
	&\quad\exists\ e: e.id = req.id \land e.v > req.id.v \\
	&\qquad\land\ \exists\ m: \texttt{allMarker}(e.\mathit{rids}, m, \textsc{CompleteRecovery}) \\
	&\qquad\qquad\ \ \land\ \texttt{allAgreed}\ (e.\mathit{rids}, m, req.id.\mathit{seqNo})
\end{aligned}
\]
\[
\begin{aligned}
&\texttt{allAgreed}(Q, m, pos) \triangleq \forall\ 1 \leq \mathit{sn} \leq pos: \\
&\quad \exists\ e:\ \forall r \in Q:\ e \in pv_r \land e.v = m.v \land e.id = (m.v, \mathit{sn})
\end{aligned}
\]

Different from Case 1, if $req$ is committed during recovery, then the Corfu's recovery protocol ensures that both $req$ and the requests prior to its position (i.e., the prefix of $req$) must have been fully replicated in the current view, and the type of each request is commonly agreed by all replicas, which is defined as $\texttt{allAgreed}(Q, m, pos)$.

Both cases reflect Corfu's replication semantics: committing a request requires every replica in the chain rather than a majority. To determine the fate of in-flight requests, the proof peeks into later views. If neither condition holds, $req$ is never committed in this execution and the ghost log is left unchanged.

\section{Prophecy Peek in LazyLog}
\label{sec:vsac-lazylog}

In LazyLog, the peek inspects the prophecy variables of all sequencers and replicas to decide whether $req$ commits and, if so, whether as \texttt{Normal} or \texttt{NOOP}. As in NOPaxos and Corfu, it distinguishes two cases.

\Para{Case 1: $req$ is committed during normal processing.} 
The proof considers $req$ committed in view $v$ once every sequencer and replica in $v$ has processed it. The leader sequencer extends the ghost log only for requests whose final position is the position $pos$ it assigned.

The peek proceeds in two steps. First, it checks that every sequencer and replica has started normal processing in view $v$, i.e., $\texttt{allMarker}(e.rids \cup e.sids, m, \textsc{StartNormal})$. 

Second, it checks that the leader in $v$ is the one responsible for $req$, i.e., $\texttt{inScope}(Q, v, pos)$. The leader assigns $req$ a local position $pos$, but that position may not be final: if the leader crashes, a leader in a later view may overwrite the position and stabilize a different order. The marker pvs settle this. $\texttt{inScope}$ holds in two cases. (1) The peek finds a \textsc{StartNormal} marker pv $m'$ of a later view $v'$ whose previous view is $v$ ($m'.\mathit{preView} = v$); $v'$ inherits $v$'s order up to $m'.\mathit{basePos}$, so if $pos < m'.\mathit{basePos}$, the position is decided by the leader in $v$ rather than by its successor. (2) If no later view ever starts normal processing, then the leader in $v$ remains the last leader and decides all subsequent requests.

\[
\begin{aligned}
&\texttt{Commit}_1(v, pos, req) \;\triangleq\; \\
&\quad\exists\ e: e.v = v \\
&\qquad\land \exists m: \texttt{allMarker}(e.rids \cup e.sids, m, \textsc{StartNormal}) \\
&\qquad\qquad\ \ \land\ \texttt{inScope} (e.rids \cup e.sids, v, pos) \\
&\qquad\qquad\ \ \land\ \texttt{allReplicated}\ (e.rids \cup e.sids, v, req)
\end{aligned}
\]

\[
\begin{aligned}
&\texttt{inScope}(Q, v, pos) \triangleq \\
&\qquad \exists r \in Q:\ \exists\ m' \in pv_r: \\
&\qquad\qquad\;  m'.\mathit{event}=\textsc{StartNormal} \\  
&\qquad\quad\land\ m'.\mathit{preView} = v \land pos< m'.\mathit{basePos} \\
&\quad \vee \neg (\exists  r \in Q:\ \exists\ m' \in pv_r: m'.\mathit{preView} = v)
\end{aligned}
\]

\[
\begin{aligned}
&\texttt{allReplicated}(Q, v, req) \triangleq \\
&\qquad \forall r \in Q:\ \exists\ e \in pv_r: \\
&\qquad\qquad e.v = v \land e.id = req.id
\end{aligned}
\]

\Para{Case 2: $req$ is committed during recovery.} This case resembles Corfu with one difference: all sequencers and replicas must have completed recovery, but only the replicas must agree on the entry pv's action (\texttt{Normal} or \texttt{NOOP}), because sequencers hold no payload and always resolve their entry pvs as \texttt{Normal}.
\[
\begin{aligned}
&\texttt{Commit}_2(v, pos, req) \;\triangleq\; \\
&\quad\exists\ e: e.id = req.id \land e.v = v \\
&\qquad\land \exists m: \\
&\qquad\quad \texttt{allMarker}(e.\mathit{rids} \cup e.\mathit{sids}, \\
&\qquad\qquad\qquad\quad m, \textsc{CompleteRecovery}) \\
&\qquad\qquad\ \ \land\ \texttt{inScope} (e.rids \cup e.sids, m.v, pos) \\
&\qquad\qquad\ \ \land\ \texttt{allAgreed}\ (e.\mathit{rids}, m, pos)
\end{aligned}
\]

\Para{Deciding committed type.} As in NOPaxos and Corfu, a committed $req$ lets the sequencer-side proof extend the ghost log atomically with an entry whose type follows $e.\mathit{action}$ in the satisfied commit condition. In LazyLog the replicas' entry pvs decide the type: if all replicas agree on $e.\mathit{action} = \texttt{Normal}$, the proof appends a \texttt{Normal} entry, meaning $req$ commits with its payload; otherwise it appends a \texttt{NOOP} entry, meaning $req$'s position is ultimately marked as a no-op.
\end{document}